\documentclass{aa}  

\usepackage{graphicx}
\usepackage{txfonts}
\usepackage{color}

\newcommand{\mps}{m\,s$^{-1}$}
\newcommand*\der{\mathop{}\!\mathrm{d}}

\begin{document}

   \title{Evolution of photospheric flows under an erupting filament in the quiet-Sun region}

   %\subtitle{}

   \author{Ji\v{r}\'i Wollmann\inst{1}
          \and
          Michal \v{S}vanda\inst{1,2}\fnmsep\thanks{Corresponding author, \email{svanda@sirrah.troja.mff.cuni.cz}}
          \and 
          David Korda\inst{1}
          \and
          Thierry Roudier\inst{3}
          }

   \institute{Charles University, Astronomical Institute, V Hole\v{s}ovi\v{c}k\'ach 2, CZ-18000 Praha 8, Czech Republic\\
         \and
             Astronomical Institute of the Czech Academy of Sciences, Fri\v{c}ova 298, CZ-25165 Ond\v{r}ejov, Czech Republic
         \and
             Institut de Recherche en Astrophysique et Plan\'etologie,   Universit\'e de Toulouse, CNRS, UPS, CNES 14 avenue Edouard Belin, F-31400 Toulouse, France
             }

   \date{Received --; accepted --}

\abstract{We studied the dynamics of the solar atmosphere in the region of a large quiet-Sun filament, which erupted on 21 October 2010. The filament eruption started at its northern end and disappeared from the H$\alpha$ line-core filtergrams line within a few hours. The very fast motions of the northern leg were recorded in ultraviolet light by the Atmospheric Imaging Assembly (AIA) imager. }
{We aim to study a wide range of available datasets describing the dynamics of the solar atmosphere for five days around the filament eruption. This interval covers three days of the filament evolution, one day before the filament growth and one day after the eruption. We search for possible triggers that lead to the eruption of the filament.}
{The surface velocity field in the region of the filament were measured by means of time--distance helioseismology and coherent structure tracking. The apparent velocities in the higher atmosphere were estimated by tracking the features in the 30.4~nm AIA observations. To capture the evolution of the magnetic field, we extrapolated the photospheric line-of-sight magnetograms and also computed the decay index of the magnetic field.}
{We found that photospheric velocity fields showed some peculiarities. Before the filament activation, we observed a temporal increase of the converging flows towards the filament's spine. In addition, the mean squared velocity increased temporarily before the activation and peaked just before it, followed by a steep decrease. We further see an increase in the average shear of the zonal flow component in the filament's region, followed by a steep decrease. The photospheric line-of-sight magnetic field shows a persistent increase of induction eastward from the filament spine. The decay index of the magnetic field at heights around 10~Mm shows a value larger than critical one at the connecting point of the northern filament end. The value of the decay index increases monotonically there until the filament activation. Then, it decreased sharply. }
{} 
% 5 {} token are mandatory

   \keywords{Sun: filaments, prominences -- Sun: atmosphere -- Sun: magnetic fields}

   \maketitle

\section{Introduction}

The Sun constitutes a dynamical system, where magnetic and velocity fields couple. Due to the magnetic field instabilities, various forms of phenomena of solar activity emerge. Among them, prominences or filaments are very intriguing features. 

Solar filaments are large regions of very dense, cool gas, held in place by a magnetic field. They often have a loop-like shape \citep[see a review by][]{2014LRSP...11....1P}. It is commonly believed that the dense plasma is held in the dips of the magnetic arcades, where the equilibrium established between the Lorentz and gravity forces \citep{1957ZA.....43...36K,2010SSRv..151..333M}. The stability of the filament then entirely depends on the stability or lability of this equilibrium. Filaments are embedded in the vicinity of the neutral line, and their legs are rooted in the chromosphere \citep{2013ApJ...777..108S,2013SoPh..282..147L}. Fibrils in the chromosphere around the polarity inversion line form a specific channel, which is arranged along the filament axis \citep{1998SoPh..182..107M} contrary to fibrils in the undisturbed chromosphere. The channel type filament body consists of a spine, barbs, and two ends. The spine usually has an inner fine structure, which contains threads that are oriented predominantly in the spinal direction of the filament \citep{1998A&A...329.1125A}.

%\subsection{Flows affecting filament stability}
The solar atmosphere, starting from the photosphere through the chromosphere and all the way towards the corona, is not static. The gas dynamics in the atmosphere and also below, in the upper layers of the solar convection zone, influence the stability of the structures of the magnetic field, thereby leading to the evolution of these structures. In the photosphere, magnetic fields are subject to diffusion due to supergranular flows and the large-scale motions of differential rotation and meridional circulation. The magnetic field elements that are transported across the solar surface can be sheared by dynamic near-surface motions, which in turn causes the shearing of the chromospheric and coronal magnetic field. This process corresponds to the formation of coronal flux ropes \citep{Mackay_Gaizauskas_2003, Mackay_Ballegooijen_2006b}. It is believed that flux ropes are important ingredients of the filaments' configuration.  During their existence, filaments undergo both quiescent and active periods.  Numerical models show that the magnetic flux rope involved in solar filament formation may be stable for many days and then suddenly become unstable, resulting in a filament eruption.  The mechanisms driving the filament eruption and disappearance are still uncertain.

The activation of a quiescent filament may be triggered by the emergence of the new magnetic flux
near a neutral line \citep{1995JGR...100.3355F,2010SSRv..151..333M}. Other studies suggest that also the filament destabilisation may also be connected to oscillations \citep{Pouget+_2006,2003ApJ...584L.103J,2012ApJ...750L...1L,2011JGRA..116.4108S,2013ApJ...773...93S}.

\begin{figure*}
    \centering
    \includegraphics[width=0.49\textwidth]{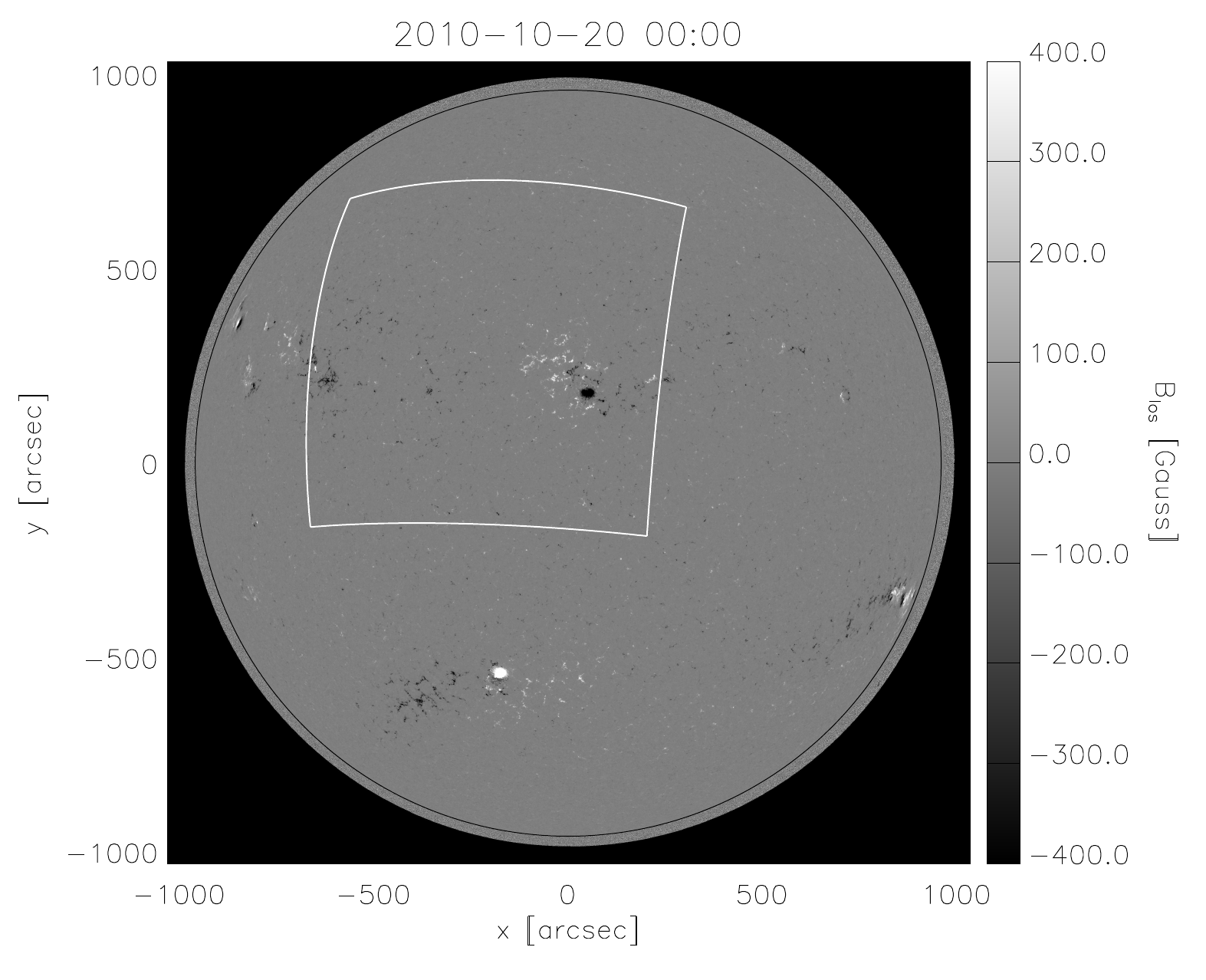}
    \includegraphics[width=0.49\textwidth]{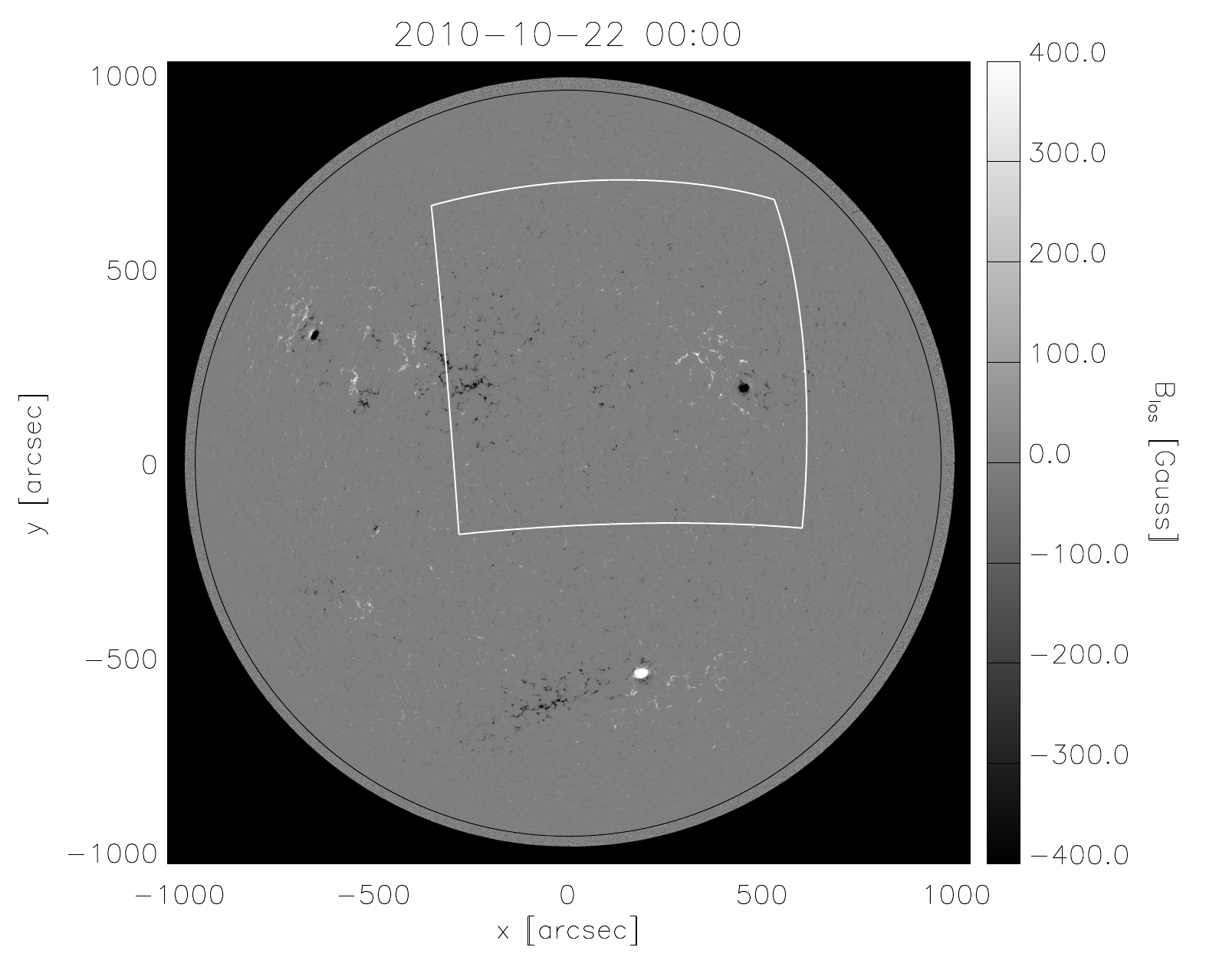}
\caption{Context images of the line-of-sight magnetic field from SDO/HMI before (left) and after (right) the filament eruption. The black line indicates the true position of the solar limb (as indicated by keywords in FITS headers), the white lines indicate the boundaries of the region of interest. }
    \label{fig:context}
\end{figure*}

\begin{figure*}
    \centering
    \includegraphics[width=\textwidth]{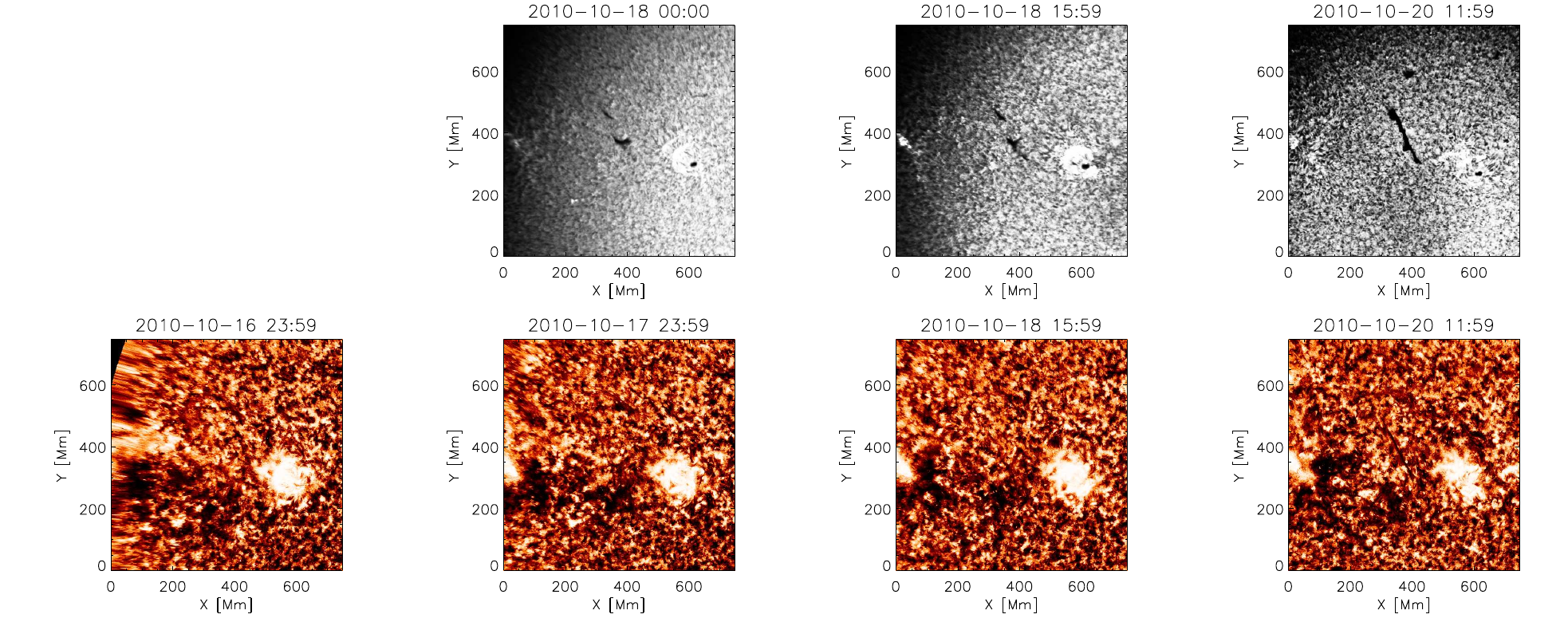}
    \includegraphics[width=\textwidth]{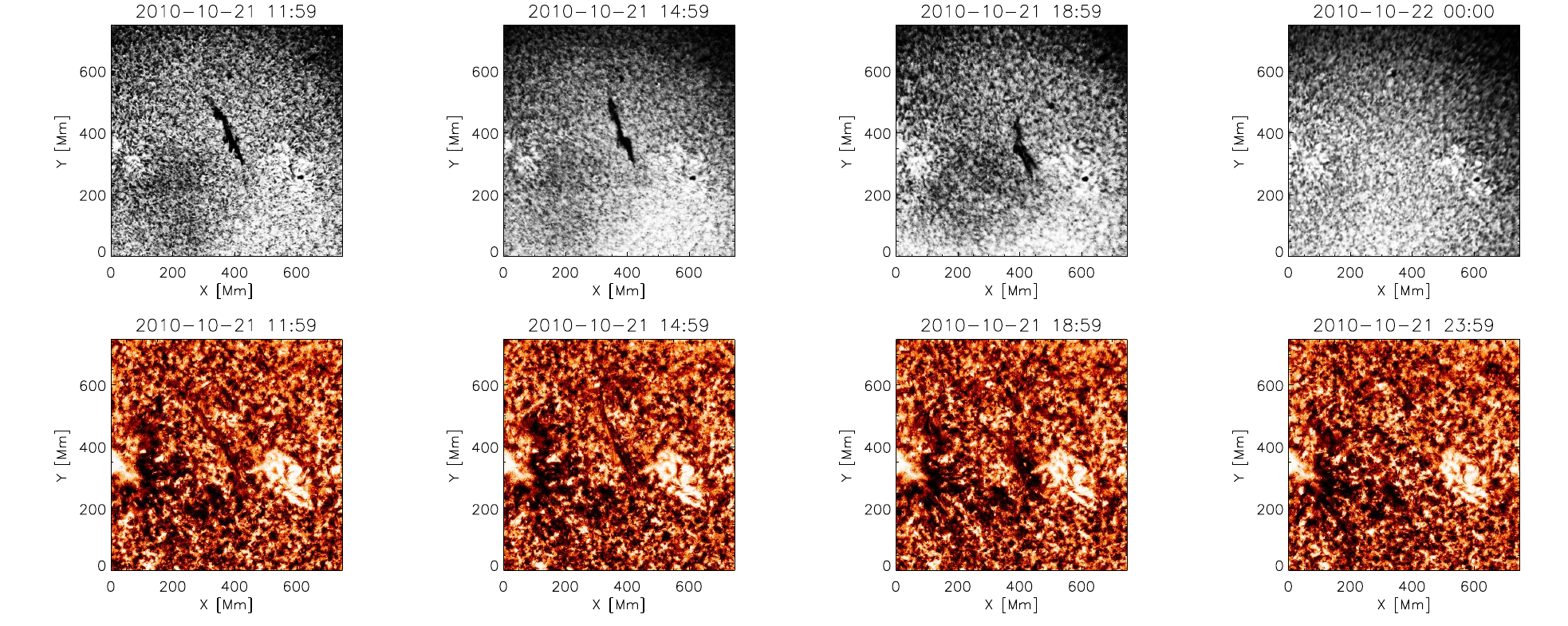}
\caption{Evolution of the filament's field of view in H$\alpha$ filtergrams (greyish panels) in comparison with corresponding views in AIA 30.4~nm filtergrams (orangish panels). A corresponding H$\alpha$ image for the first frame is not available, thus the panel is empty. We note that for display purposes, a histogram equalisation was used to improve the visibility of structures.}
    \label{fig:evolution}
\end{figure*}

Studies also point out the importance of peculiarities in the photospheric flows. For instance, \cite{Rondi+_2007} measured horizontal velocities in the vicinity of and beneath a filament before and during the filament’s eruptive phases by tracking granules in white-light images. They showed that both parasitic and normal magnetic polarities were continuously swept into the filament gap by the diverging supergranular flow. They also observed the interaction of opposite polarities in the same region, which could be a reason for initiating the destabilisation of the filament by triggering a reorganisation of the magnetic field. \cite{2008A&A...480..255R} showed peculiar velocity features in the photosphere during the eruption of a quiescent filament. The horizontal velocities were determined by tracking doppler features in full-disc dopplergrams and by tracking magnetic elements in full-disc magnetograms. They found evidence for a parallel flow along the filament prior to its eruption, which disappeared after the filament vanished. They also found that the shear in zonal velocities near the ignition point of the filament eruption increased monotonically before the eruption and dropped suddenly immediately after it. They concluded that the shear pumped the energy to magnetic structures thereby triggering the instability and eruption of the filament. \cite{2014A&A...564A.104S} computed the vector velocity maps by tracking granules in intensity maps for a single filament using different spatial resolutions. The velocity field did not show the presence of large-scale flows. They found that the diverging flows inside the supergranules were similar in and out of the filament channel. The converging flows were identified around the filament footpoints and at the edges of the filament channel. \cite{2006ApJ...653..725H} used helioseismic inversions to determine velocity fields in the vicinity of the filament over four days. A time-averaged velocity field showed signs of a significant shear current parallel to the polarity inversion line. \cite{2018A&A...613A..39A} performed a statistical study of 64 quiescent filaments, for which they studied velocity fields obtained by tracking supergranules represented in dopplergrams in their vicinity. For a large fraction of the sample, they indeed found evidence for shearing motions along the filament spines. On the other hand, they did not find compelling evidence for converging motions towards polarity inversion lines. They also pointed out that filament barbs are mostly anchored in the places with significantly convergent flows. Perhaps a dispersed magnetic flux concentrates here giving rise to a new magnetic-field structure that connects with the filament with a barb later. 

When the filament activates, various phenomena may be observed. \cite{2009SoPh..259...13G} investigated the disappearance of filaments at various levels in the solar atmosphere using both ground-based and space-bourne data. They found that vortical motions below the filament were observed a day before the filament’s disappearance. The authors suggested that the photospheric shear motions around lateral barbs caused magnetic flux cancellation and led to the destabilisation of the magnetic system in the filament. The filament then expanded and gradually disappeared.

%\subsection{Studied Filament}
\section{Observations}
A large filament was observed east from active regions NOAA~11113 in October 2010. The filament was visible above the eastern solar limb between latitudes of 10 and 30 degrees on 13 October in Atmospheric Imaging Assembly \citep[AIA;][]{2012SoPh..275...17L,2012SoPh..275...41B} 30.4~nm filter and it moved towards the central solar meridian in the following week. Its visibility was increasing gradually in AIA 30.4~nm until 21 October, probably due to the increasing opacity caused by filling the filament with plasma. The filament disappeared suddenly in the afternoon of 21 October and was not visible any longer in the following days. 

According to the Helioseismic and Magnetic Imager \citep[HMI;][]{2012SoPh..275..327S, 2012SoPh..275..207S} line-of-sight context magnetograms (see Fig.~\ref{fig:context}), the investigated filament was placed roughly above a polarity inversion line (PIL) separating the trailing polarity (positive) of NOAA~11113 and the dispersed (its negative part) magnetic field located west of active region NOAA~11118. This dispersed negative field was probably the leftover of an active region that dispersed there before. Within the negative part of this dispersed field, a new active region NOAA~11119 emerged on 24 October. 

In the Global Oscillations Network Group \citep[GONG;][]{1996Sci...272.1284H} H$\alpha$ context images (line-core filtergrams) the filament extended roughly between latitudes 18 and 33 degrees north of the solar equator. Until 19 October the filament evolved slowly, then during the days of 20 and 21 October, it grew significantly in area, visible as absorption in the H$\alpha$ line, and dropped suddenly within a few hours in the afternoon of 21 October.

\subsection{Analysed datasets}
The evolution of the filament was captured by instruments both in space, namely with the HMI and AIA aboard the Solar Dynamics Observatory \citep[SDO;][]{2012SoPh..275....3P},  and on the Earth. In our study, we mainly used the following data series:

\begin{description}
\item[SDO/HMI dopplergrams:] The 45-second cadence full-disc dopplergrams were used as a primary source for measuring the travel times of helioseismic waves to measure the near-surface flow field. We studied, in total, five days around the filament disappearance, that is from 18 to 22 October. 
\item[SDO/HMI magnetograms:] Line-of-sight magnetograms measured by SDO/HMI were used firstly to check the context of the field of view and to discuss the possible connectivity of the filament. In the following, the magnetograms at a reduced cadence of one hour were obtained by averaging available frames over one hour. They were used in the linear extrapolation of the magnetic field to study the structures of the potential magnetic field components.
\item[SDO/HMI intensitygrams and dopplergrams:] For the day of the eruption, we tracked the granules visible in the HMI Intensitygrams to obtain a high-resolution high-cadence horizontal velocity field utilising coherent structure tracking \citep[CST;][]{2007A&A...471..687R,2017A&A...599A..69R}. The CST velocities are complemented by HMI dopplergrams to form full velocity vectors. The pipeline running at Institut de Recherche en Astrophysique \& Plan\'{e}tologie (IRAP) resulted in full-disc maps of the zonal, meridional, and radial components of the surface flow with a cadence of 30~min. 
\item[GONG H$\alpha$ filtergrams:] The full-cadence filtergrams measured in the H$\alpha$ line by the ground-based GONG network were used mainly to study the filament evolution and to determine the various phases of the filament dynamical behaviour. 
\item[Context data:] We used several other contextual data (such as the already mentioned SDO/AIA 30.4~nm filtergrams) to follow the filament evolution. 
\end{description}

\noindent All SDO datasets were processed using standard JSOC web-based interface. That included the tracking of the patch of the solar surface projected to Postel coordinates. The patch was centred at Carrington coordinates of $l=122^\circ$ and $b=25^\circ$ with a pixel size of $0.12^\circ$ and a square field of view of $512\times512$~pixel. Similar tracking with the same projection and pixel size was performed using our IDL code for ground-based GONG H$\alpha$ data. Using this procedure, we automatically ensured the mutual co-alignment of all the series. 

\subsection{Velocity fields}
To study the dynamics of the photosphere in the vicinity of the filament, we further used vector flow maps derived by time--distance helioseismology. By inverting the travel times of the waves, we inferred estimates of components of plasma flow near the surface in a pseudo-Cartesian coordinate system $(X,Y,Z)$, where $X$ indicates the horizontal east--west direction along solar equator, $Y$ the horizontal south-north meridional direction and $Z$ indicates the vertical (radial) coordinate. The methodology is thoroughly described elsewhere, the maps we have at our disposal are principally (by methodology) analogous to the maps carefully tested by \cite{2013ApJ...771...32S}. 

We focused on the analysis of the maps of three components of the vector flows co-spatial with H$\alpha$ and AIA 30.4~nm frames. These maps have an effective resolution of about 10~Mm and are averaged over 6, 12, and 24 hours with a critical sampling. The levels of random noises in the maps are 18~\mps{} for the horizontal components, and 3~\mps{} for the vertical component, respectively, when considering averaging over 24 hours. The noise estimates strictly apply only to quiet-Sun regions and are based on statistical properties of solar oscillations. Our region of interest near the filament shows only weak dispersed magnetic field, thus, noise estimates there are also representative here. For the shorter time averaging, the random-noise levels scale as $\sqrt{T}$, where $T$ is the length of averaging. 

In the flow maps, the structure of supergranules is clearly visible. The divergent centres of supergranular cells coincide with upflows in the maps of vertical velocity. In the following, we ignore the distortions caused by Postel projection, because we focus on the region close to the centre of the map, where the distortions are negligible. The differentiable nature of the flow maps allowed us to compute the flow field derivatives, namely the horizontal divergence $\nabla_h \cdot \boldsymbol{v}_h,$ and the vertical component of vorticity $(\boldsymbol{\nabla}\times \boldsymbol{v})_Z$ defined by
\begin{equation}
    \nabla_h \cdot \boldsymbol{v}_h = \frac{\partial v_X}{\partial X}+\frac{\partial v_Y}{\partial Y}\quad {\rm and}\quad (\boldsymbol{\nabla}\times \boldsymbol{v})_Z=\frac{\partial v_Y}{\partial X}-\frac{\partial v_X}{\partial Y}.
    \label{eq:derivatives}
\end{equation}

%\section{Filament Eruption}
\section{Results\\ }
\begin{figure}
    \centering
    \includegraphics[width=0.5\textwidth]{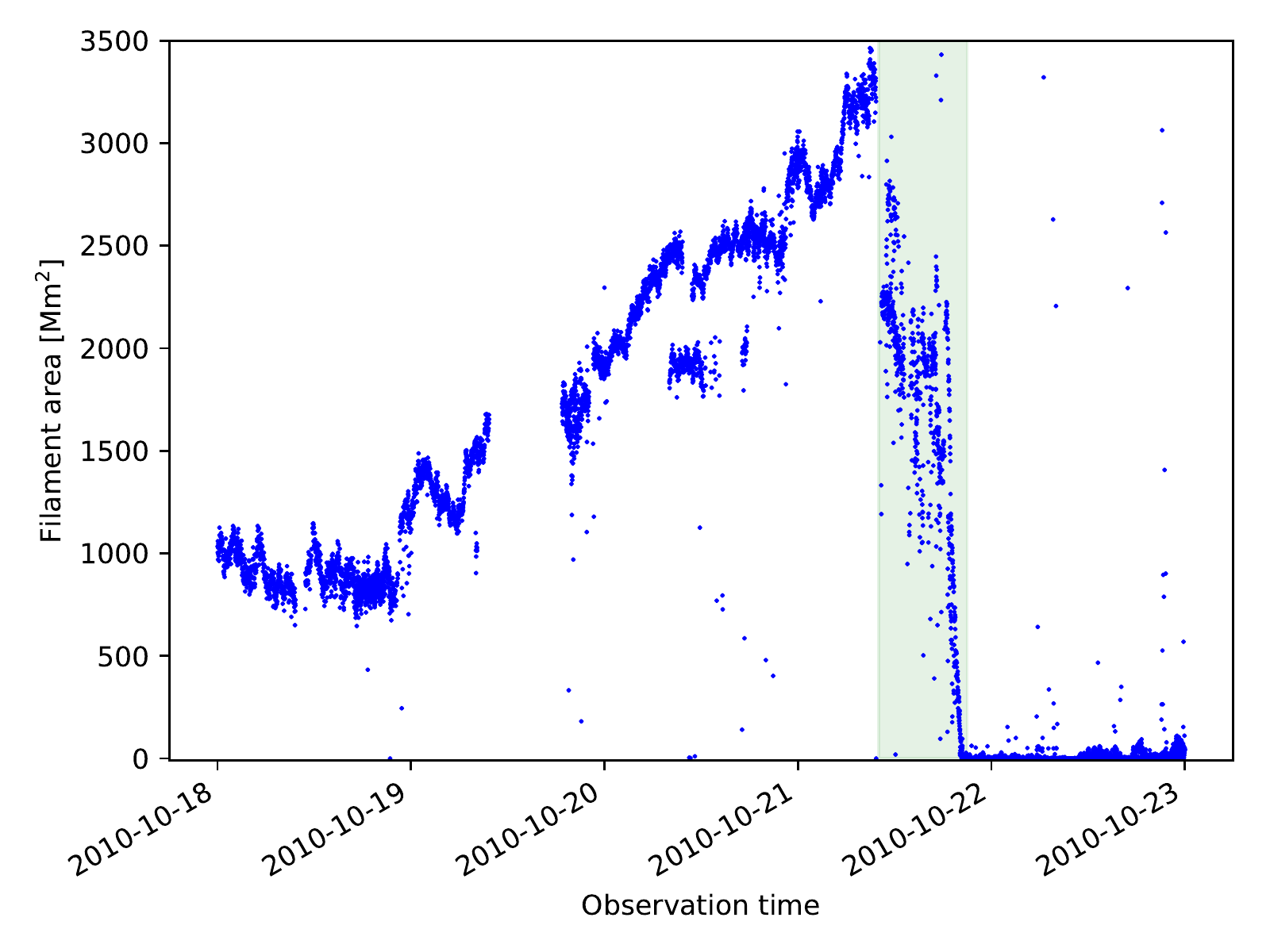}
    \caption{Evolution of filament area in H$\alpha$ filtergram in time. The greenish box indicates the activation and eruptive phases of the filament's life. }
    \label{fig:life}
\end{figure}
Various observations allowed us to describe particular aspects of the process of filament destabilisation. 

\subsection{Filament in H$\alpha$}
In order to study the changes in the photosphere underlying the filament, we first need to identify various phases of the filament's life. In order to do so, we mainly used a 5-day series of GONG H$\alpha$ images. The sequence was processed as described in the previous section. The tracking ensured that the datacube remained centred on the position of the filament (see examples in Fig.~\ref{fig:evolution}). 

The filament pixels were identified by the thresholding method. The threshold was empirically set to $0.92 I_{\rm mean}$, where $I_{\rm mean}$ represents the mean intensity of all pixels in the datacube. We then measured the total area taken by the filament in the frames for each frame separately. When measuring the area in physical units we neglected the distortions caused by the Postel projection, because the filament was very close to the centre of the field of view. 

The evolution of the area is plotted in Fig.~\ref{fig:life}. We plot all the measurements even in the case when the measured area is an obvious outlier. These points have an origin mostly in frames when the solar disc was not entirely clear of clouds. 

We see that at the beginning, the area of the filament remained almost constant at a value of about 1000~Mm$^2$. At the end of the day on 18 October, the area started to grow and continued to do so until midday of 21 October. This is the growth phase of the filament. Plasma is fed into the structures of the filament's magnetic field. 

At around 11:00 UT on 21 October, the area started to decrease because of filament activation. The plasma is no longer fed to the filament's magnetic field and slowly disappears. In the online movie, the shape of the filament starts to change. This is the activation phase of the filament.

Finally, between about 19:00 UT and 20:15 UT, the filament rapidly vanished from the H$\alpha$ filtergrams. The filament erupted and disappeared. 

\subsection{Evolution in the ultraviolet images}
\begin{figure}
    \centering
    \includegraphics[width=0.45\textwidth]{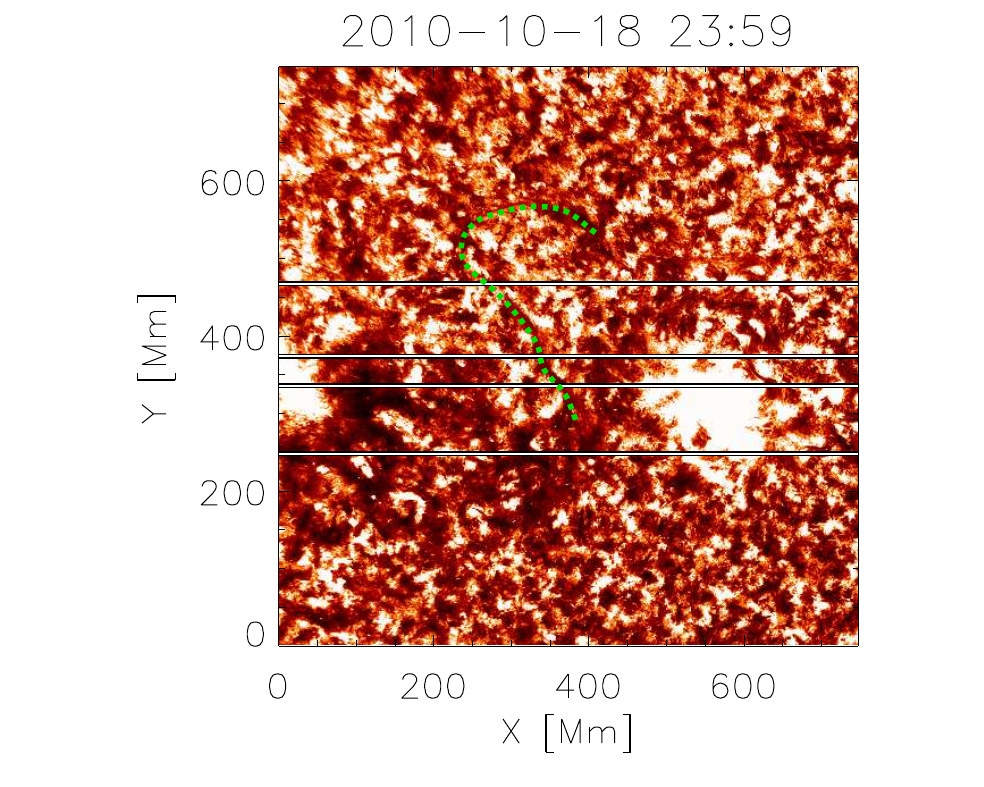}
\caption{Field of view in the 30.4~nm spectral band to indicate the positions of the horizontal cuts, in which the time--distance diagrams are plotted in Fig.~\ref{fig:T-D304}. With a green dotted line, the position of the hook connecting the filament to a northern footpoint is outlined, together with the spine of the filament.}
    \label{fig:cuts304}
\end{figure}

\begin{figure*}
    \centering
    \includegraphics[width=\textwidth]{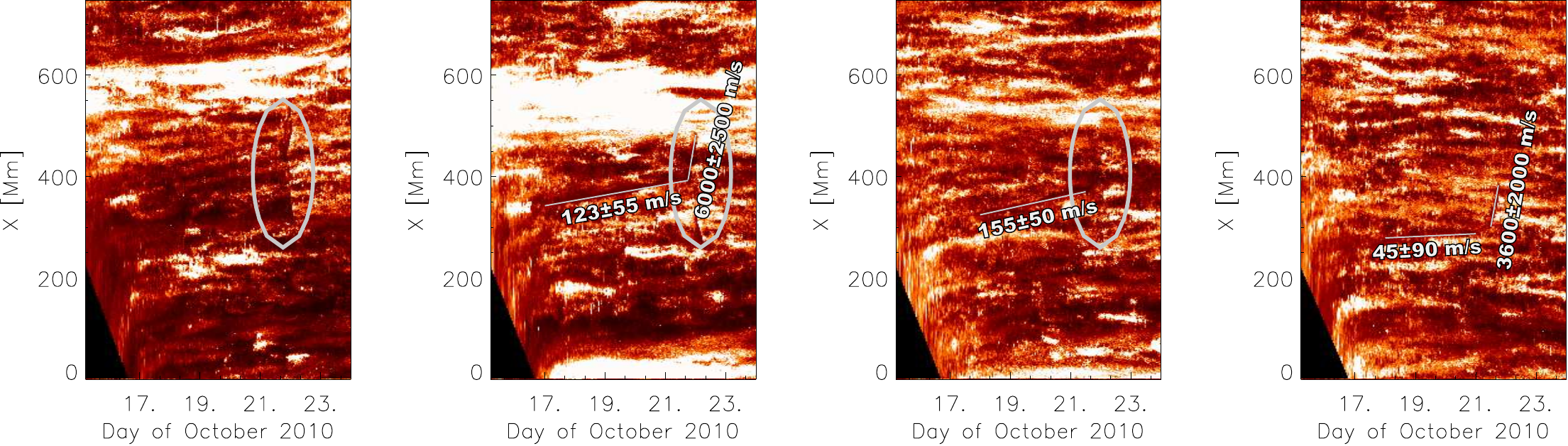}
\caption{Time--distance diagrams for the cuts indicated in Fig.~\ref{fig:cuts304}. In the left-most panel, the transition from the darker filament to the brighter band is marked by an ellipse. In the remaining panels we indicate the position of the filament and add also an estimate of the velocity, with which the structure propagates along the horizontal axis.}
    \label{fig:T-D304}
\end{figure*}

In the online movie of AIA 30.4~nm (attached online) frames tracked with the rate of the Carrington rotation the absorbing plasma in the filament was first clearly visible on 17 October at roughly 04:00~UT as a very narrow linear structure stretching diagonally through the field of view. Its thickness started to grow rapidly on 18 October at roughly 15:00~UT, a few hours earlier than the growth of the area of the absorption in H$\alpha$ was observed. 

Throughout the whole period, the spine of the filament in AIA 30.4~nm moved westwards with respect to the Carrington rotation with a speed of about 100~\mps{}, as if the whole structure was pushed away by the monotonic growth of dispersed magnetic field west of NOAA~11118 (see Figs.~\ref{fig:cuts304} and~\ref{fig:T-D304}) and by differential rotation. 

From the AIA 30.4~nm movie, it is also visible that the whole diagonal filament is connected to the footpoint north from the filament spine by a large hook. This hook is barely visible in still frames, however, it is recognisable in the online time-lapse movies. The position of the hook is indicated in Fig.~\ref{fig:cuts304}. 

A fast evolution at the northern and southern filament footpoints starts to be observed on 21 October at around 12:00~UT. The filament ends, visible in H$\alpha$ at both sides, are moving very fast (at speeds of several kilometers per second) westwards and they disappear at roughly 20:00~UT on the same day. In the northern part, the filament seems to remain connected to the footpoint by the large hook, when the hook shortens and sweeps west. After the filament disappearance, especially in the lower part of the field of view in the quiet-Sun region between the magnetised regions, a quickly extending brightening band is observed parallel to the spine of the former filament. This very fast transition is also visible in Fig.~\ref{fig:T-D304}. The brightening may be either explained by a decrease in the absorption coefficient perhaps due to the decrease of the density in the arcade loops or by local heating, possibly caused by filament plasma falling down along the arcade loops. 

\subsection{Evolution of photospheric flows}
We studied the near-surface vector flow field in the vicinity of the filament evolving over five days from 18 October to 22 October. Namely, we focused on maps averaged over 6 hours with a critical sampling. Firstly, we studied a general appearance of the vector velocity field and its derivatives computed according to Eq.~\ref{eq:derivatives}, but we did not find any peculiarities indicating a trigger for the filament eruption. The dominant features over the whole field of view are the supergranular cells. In consecutive frames, most of the supergranular cells are reproduced at the same place, thereby increasing the credibility of the flow maps. 
 
We realise that useful information may be hidden in the large-amplitude local flows. In addition, a number of studies show the importance of large-scale components of the flow that appeared only after the smaller-scale components were filtered out from the full flow field \citep[e.g.][]{2008A&A...480..255R,2018A&A...613A..39A}. 

\begin{figure}
    \centering
    \includegraphics[width=0.495\textwidth]{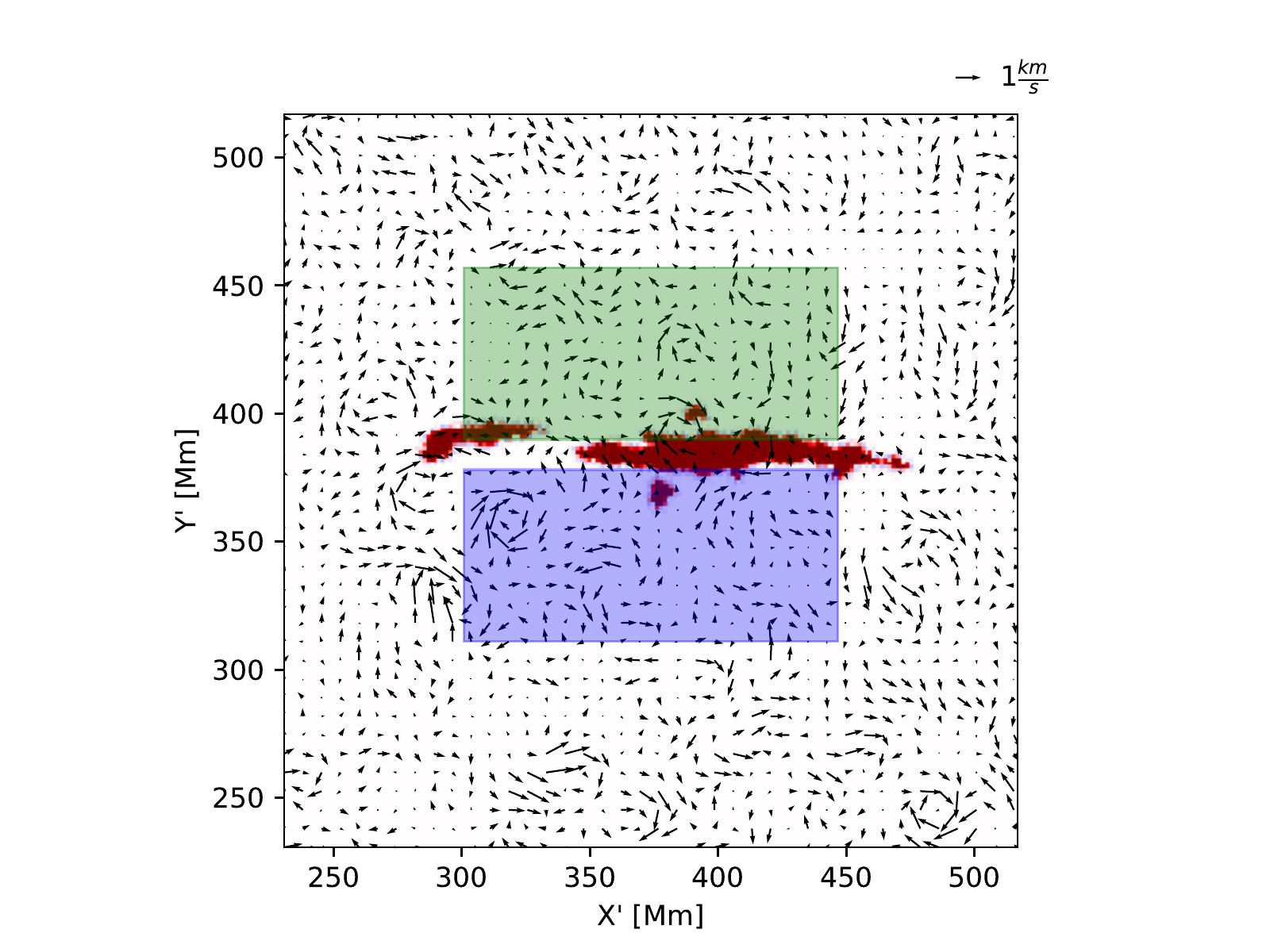}
    \includegraphics[width=0.495\textwidth]{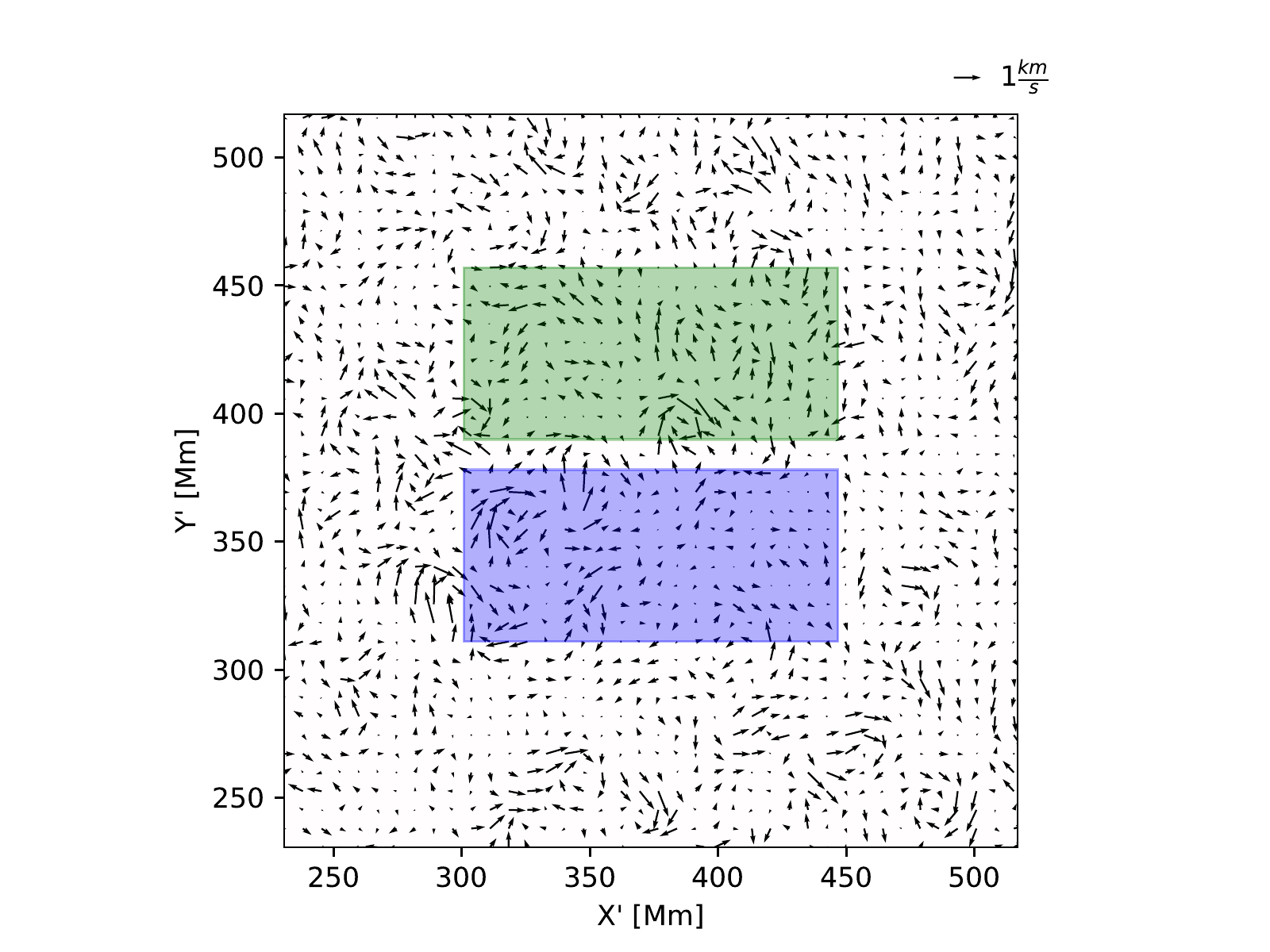}
    \caption{Filament field of view reoriented so that the filament spine is parallel to the horizontal axis. The boxes indicate the regions over which the difference of the parallel and perpendicular velocities respectively where evaluated `above' and `below' the filament. The upper panel corresponds to 15:00 UT on 21 October, the lower panel to 21:00 UT on 21 October.}
    \label{fig:rectangles}
\end{figure}

Inspired by this approach, we studied the integrated flow components in the close vicinity of the filament. Namely, we were interested in the mean flows along the filament spine $\langle v_\parallel \rangle$ and in the perpendicular direction $\langle v_\perp \rangle$. Thus, we averaged the two above mentioned components in a rectangle having a length of 150~Mm and width of 70~Mm just above and just below the spine of the filament visible in the H$\alpha$ line, as indicated in Fig.~\ref{fig:rectangles}. Then, we computed
\begin{align}
    \Delta \langle v_\parallel \rangle &= \langle v_\parallel \rangle_{\rm above} - \langle v_\parallel \rangle_{\rm below} \quad {\rm and}\\
    \Delta \langle v_\perp \rangle &= \langle v_\perp \rangle_{\rm above} - \langle v_\perp \rangle_{\rm below}. 
\end{align}
The differences of the parallel velocity components $\Delta \langle v_\parallel \rangle$ then evaluate the velocity shear along the filament axis, whereas the differences in the perpendicular velocities $\Delta \langle v_\perp \rangle$ indicate the prevalence of divergent flows from or convergent flows towards the filament axis. 

\begin{figure}
    \centering
    \includegraphics[width=0.49\textwidth]{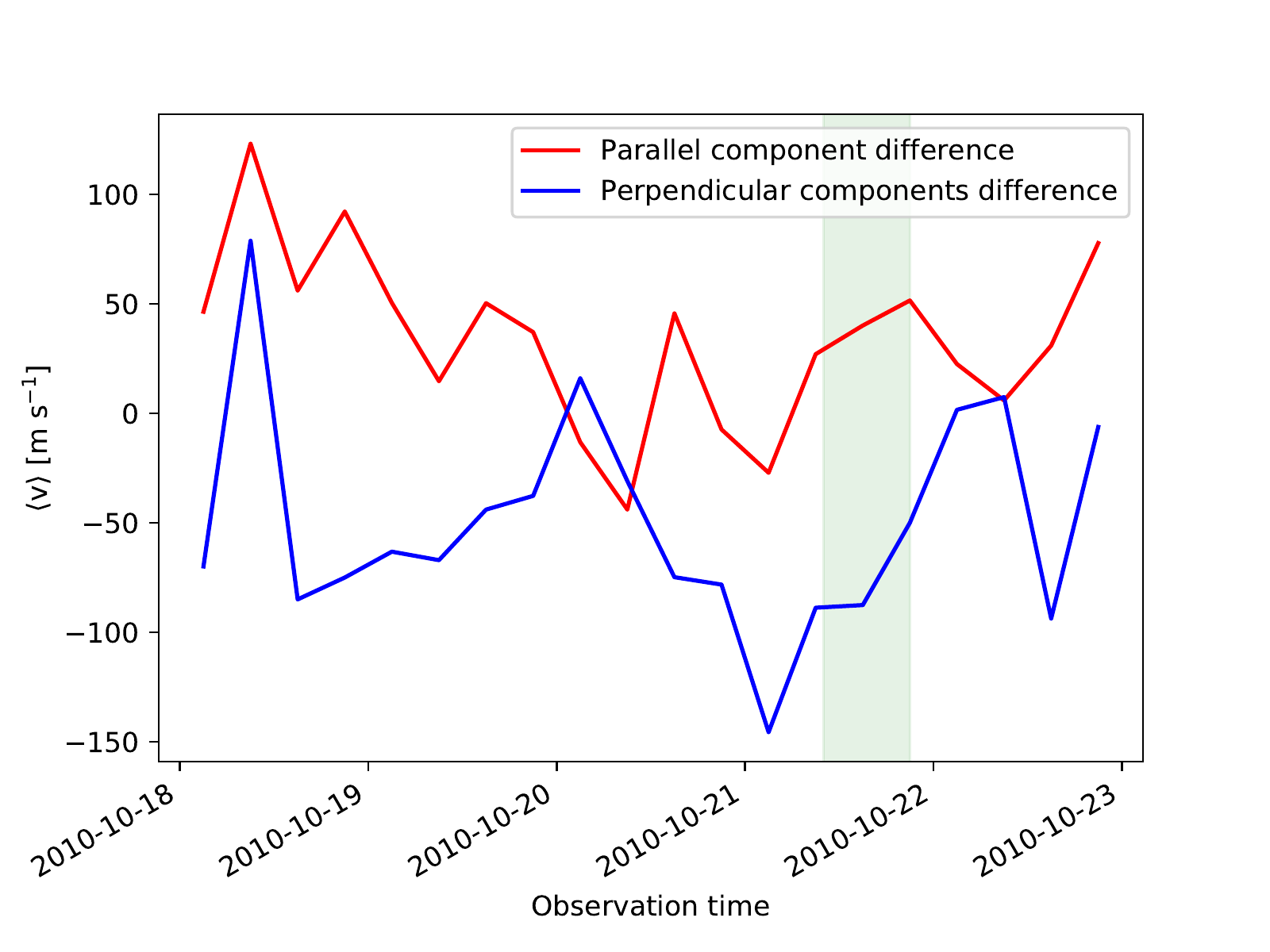}
    \includegraphics[width=0.49\textwidth]{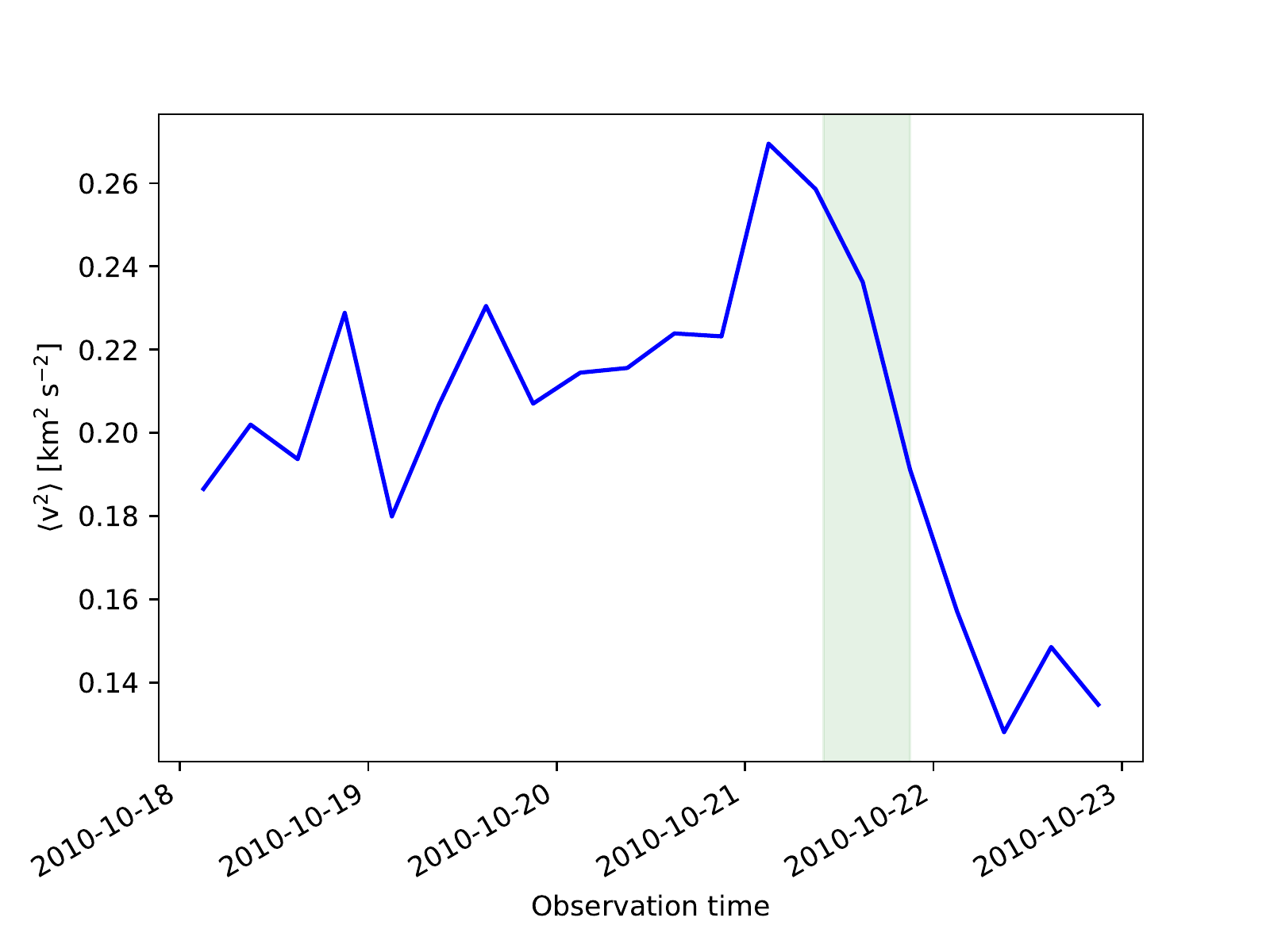}
    \caption{Upper panel: evolution of $\Delta \langle v_\parallel \rangle$ (red) and $\Delta \langle v_\perp \rangle$ (blue). Lower panel: evolution of the mean squared velocity $\langle v^2 \rangle$ in the region of the filament. The greenish bar indicates the filament activation and eruption.}
    \label{fig:differences}
\end{figure}
The time evolution of these differences is given in the left panel of Fig.~\ref{fig:differences}. The period of the filament activation and eruption is indicated by the greenish rectangle. It would seem that in the period before the filament activation the shear in the parallel components did not show a secular trend, except for the oscillations, but in the perpendicular direction, we see a trend of the growing converging (negative sign) large-scale flows towards the filament axis before the onset of the activation phase. Then, the convergent flows weakened. 

The mean squared velocity $\langle v^2 \rangle$,
\begin{equation}
    \langle v^2 \rangle= \langle v_X^2+v_Y^2+v_Z^2 \rangle_{\rm box},
    \label{eq:v_sq}
\end{equation}
averaged over the same region, also demonstrates peculiar behaviour, when we see an increase before the activation phase and a continuous decrease during the activation phase and the filament eruption (see the lower panel of Fig.~\ref{fig:differences}). This change has mostly to do with the trend in the perpendicular velocity component. 

\begin{figure*}
    \centering
    \includegraphics[width=0.3\textwidth]{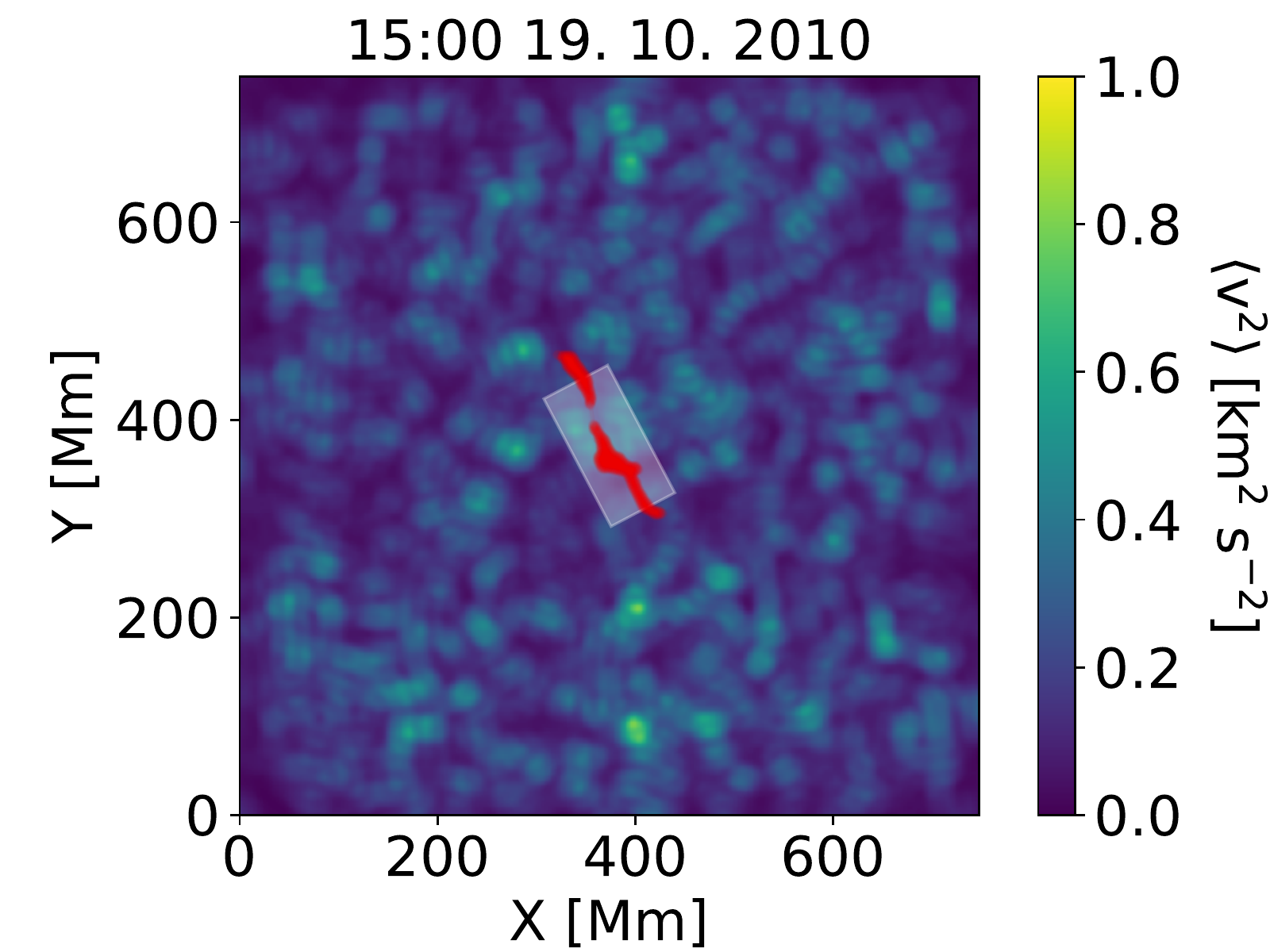}
    \includegraphics[width=0.3\textwidth]{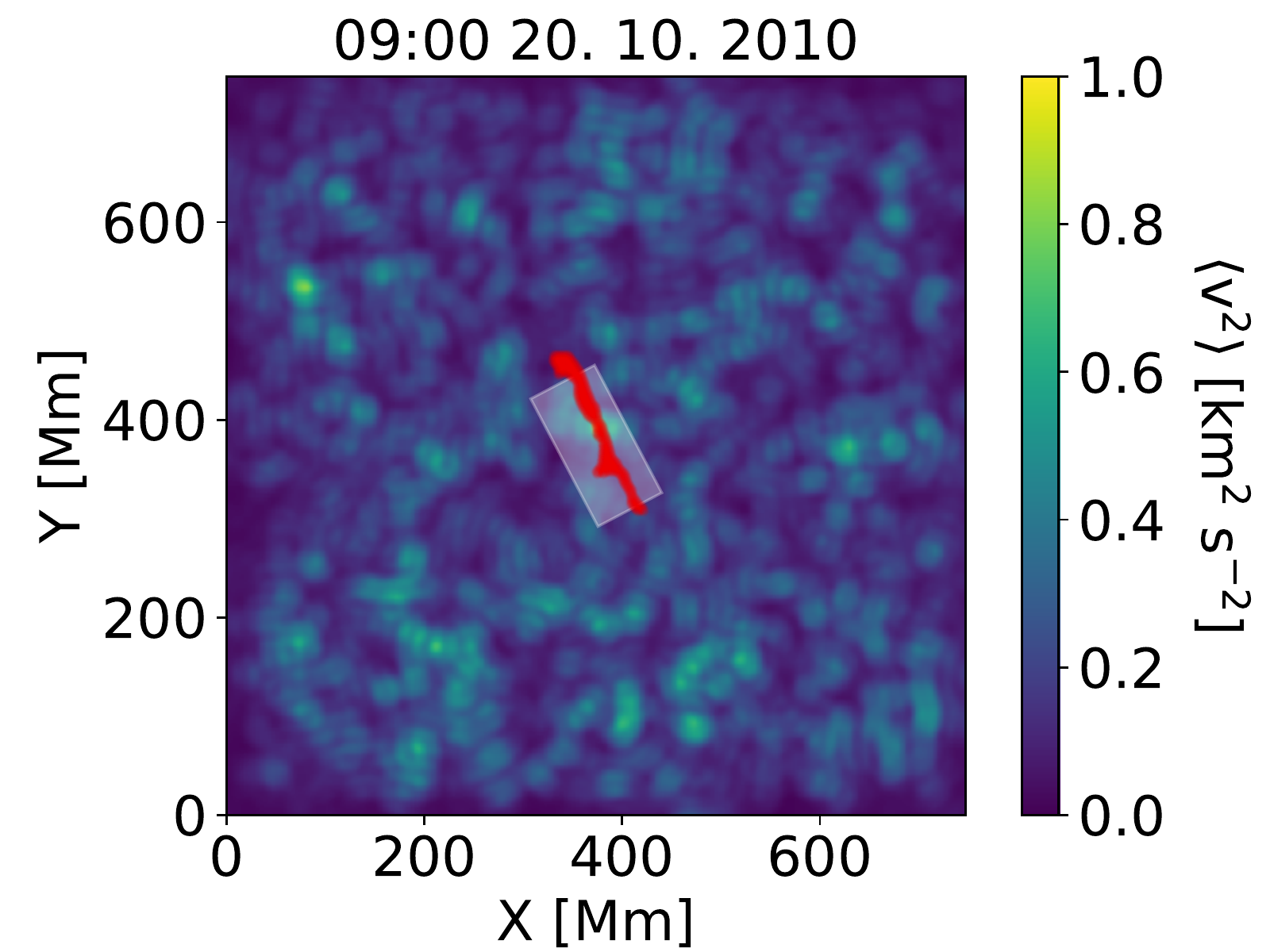}
    \includegraphics[width=0.3\textwidth]{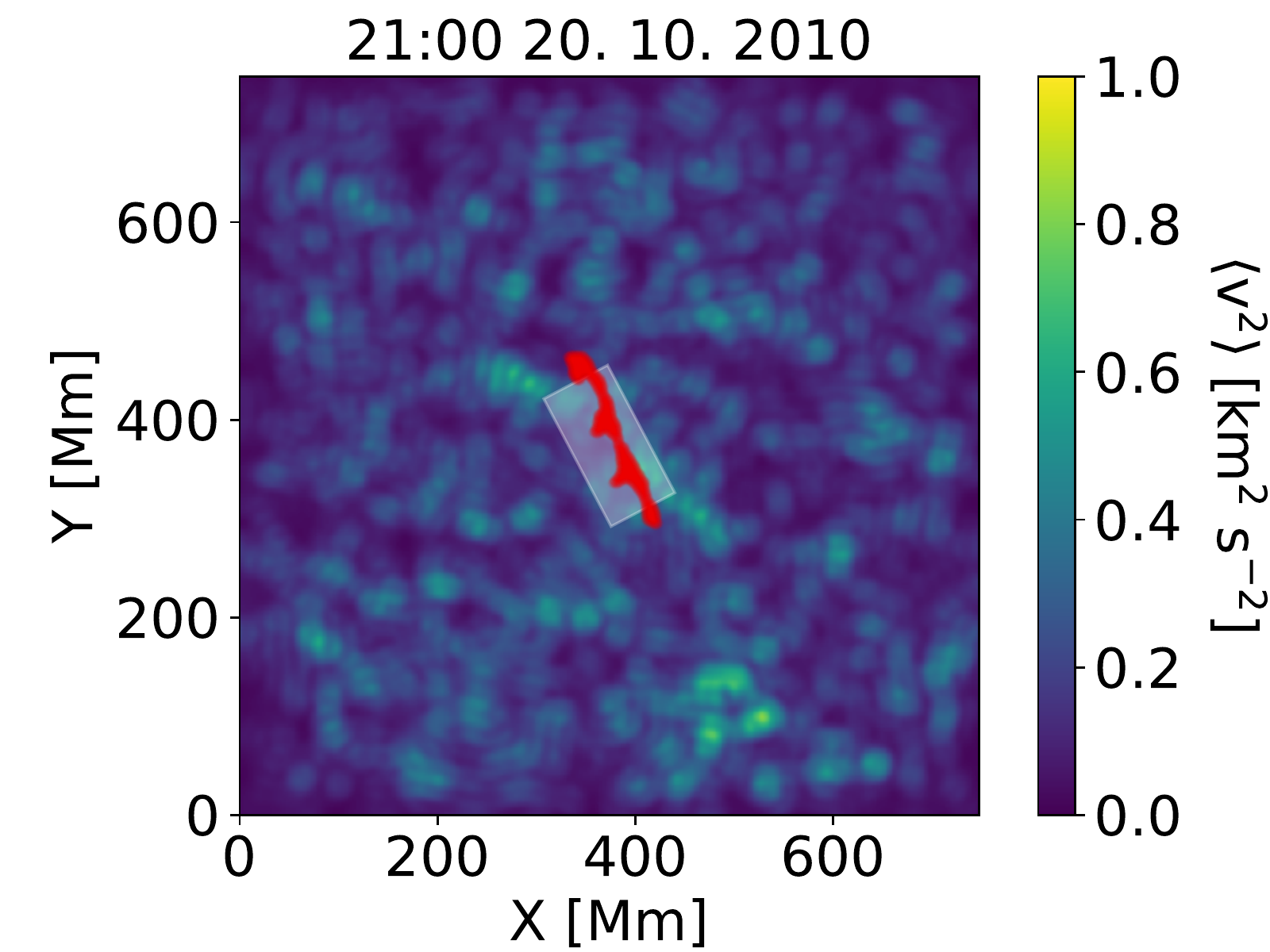}
    \includegraphics[width=0.3\textwidth]{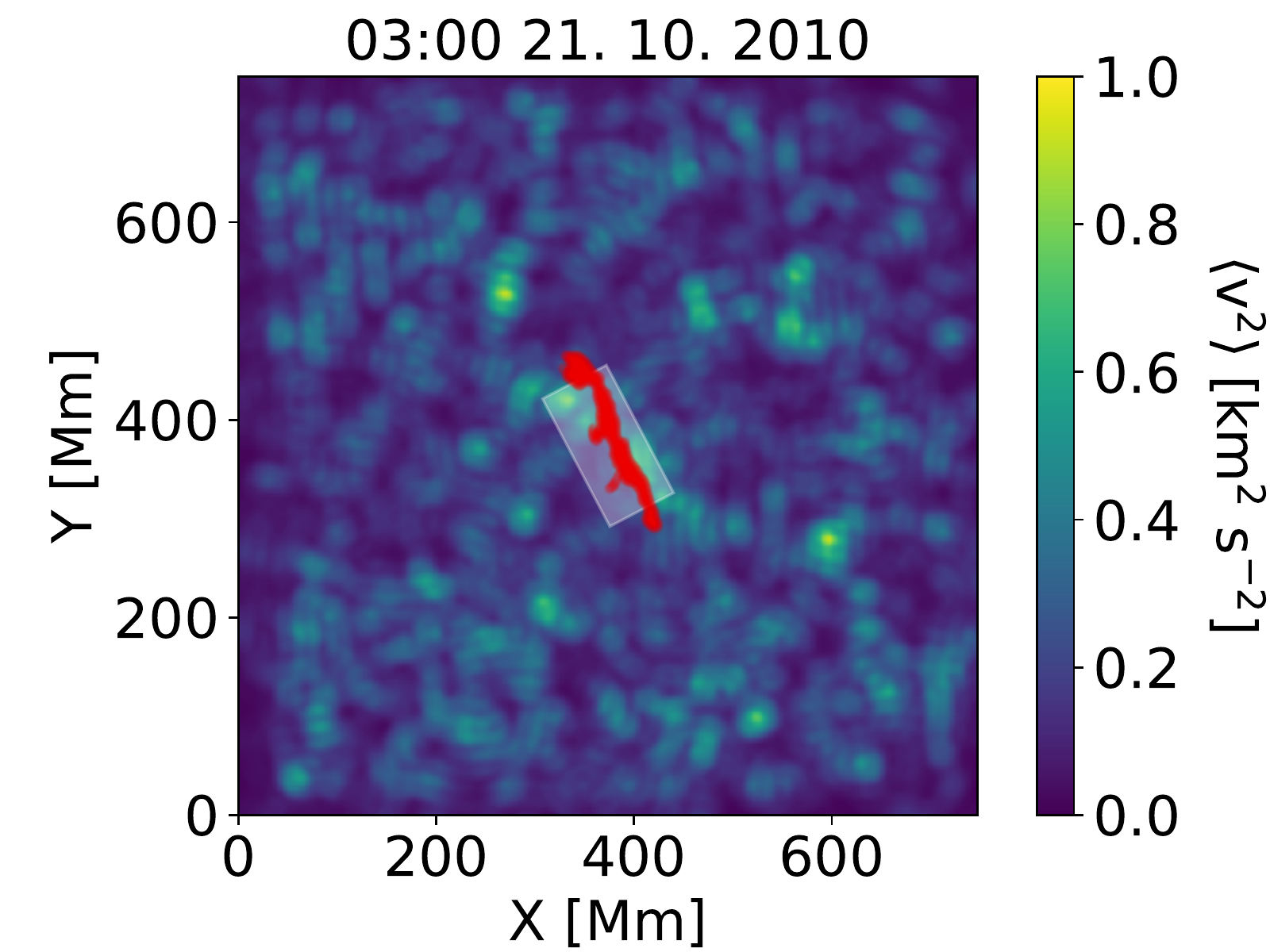}
    \includegraphics[width=0.3\textwidth]{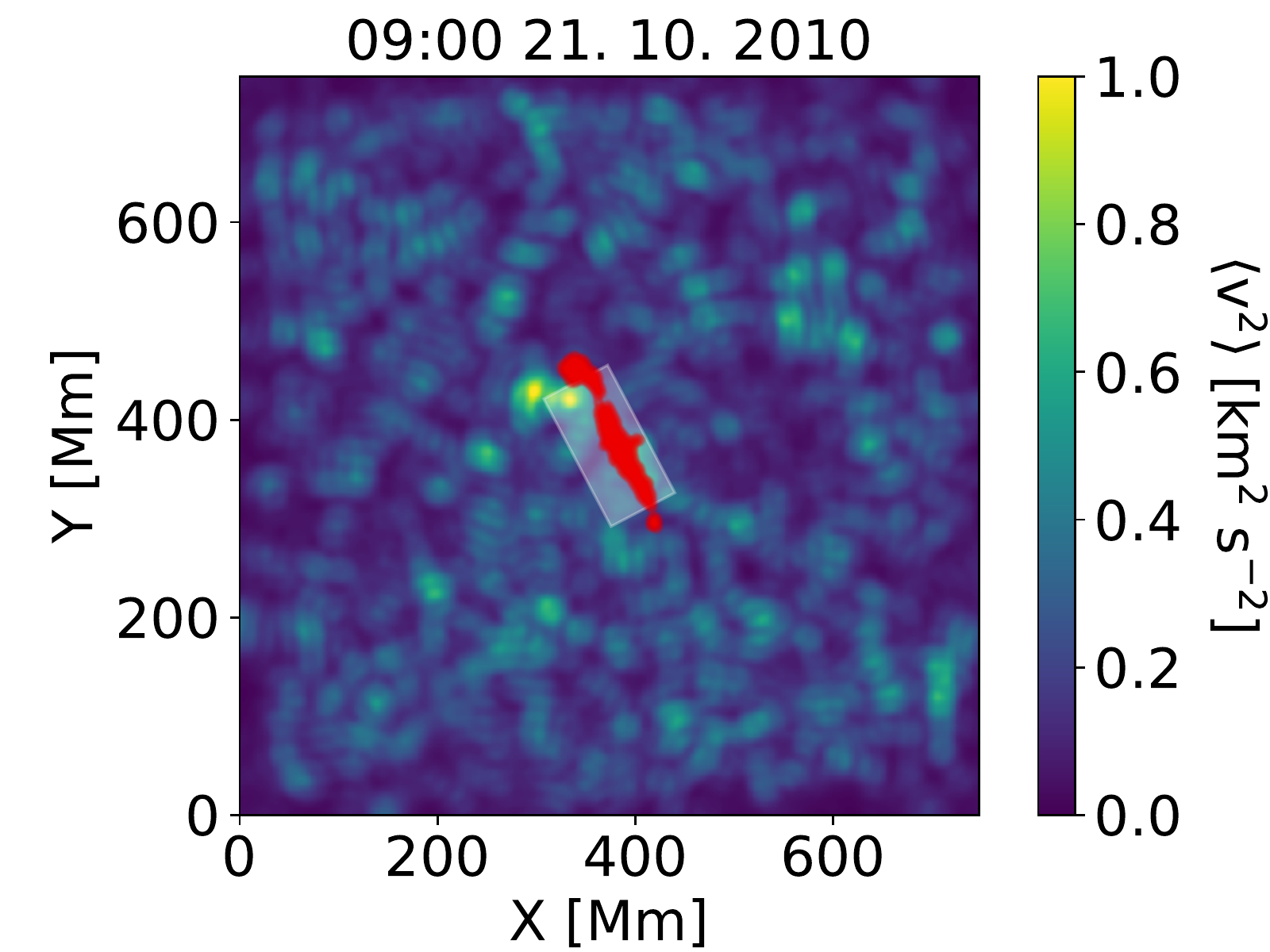}
    \includegraphics[width=0.3\textwidth]{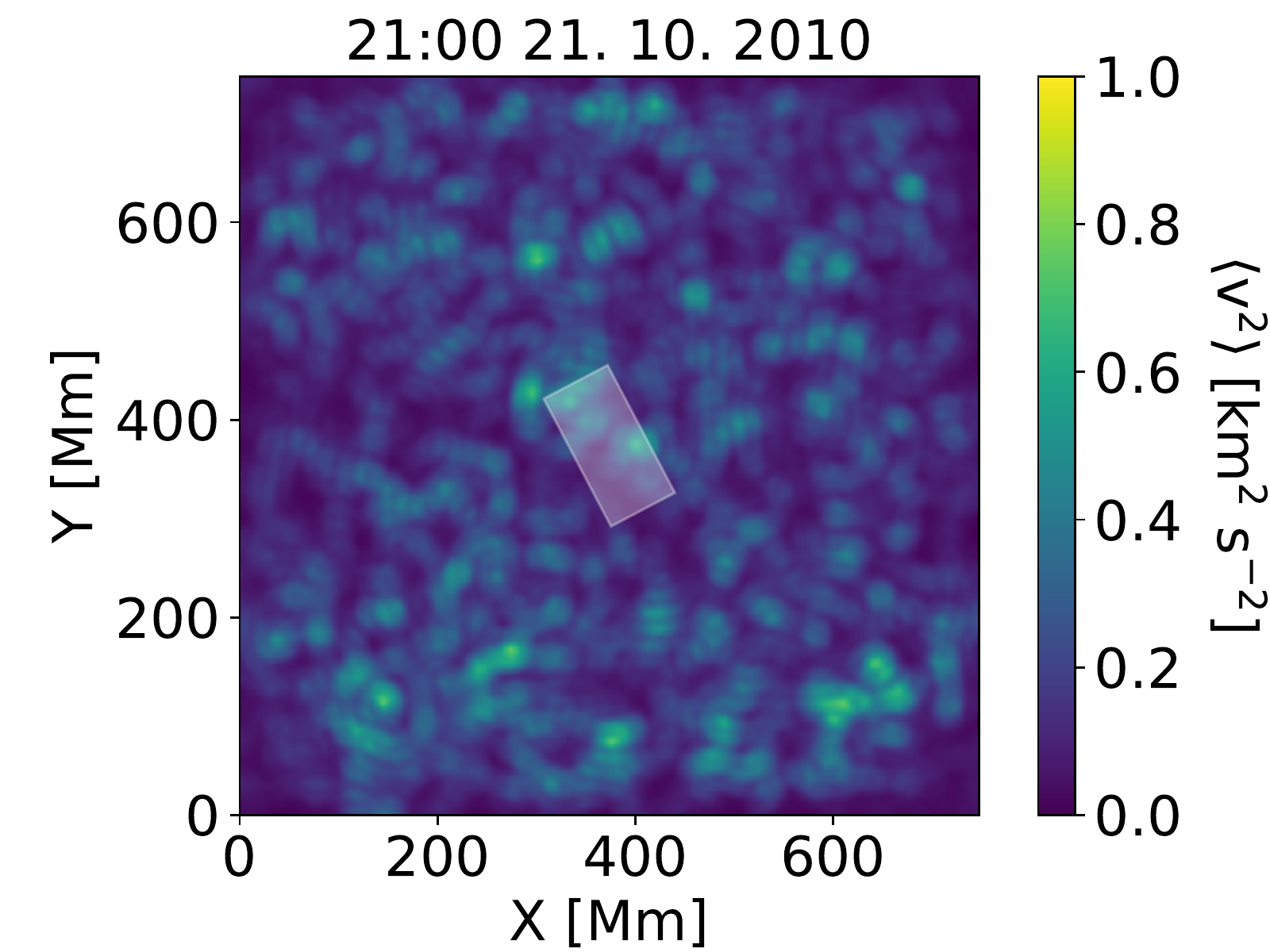}
    \caption{Maps of mean squared velocity in the region of the filament. The corresponding shape of the filament is overplotted in red. The semitransparent rectangle represents the area over which the square velocity was computed to be plotted in Fig.~\ref{fig:differences} bottom panel.}
    \label{fig:horvel}
\end{figure*}

We studied the distribution of the mean squared velocity over the field of view.  Thus, we computed $\langle v^2 \rangle$ according to (\ref{eq:v_sq}) in a sliding window having a radius of 15~Mm. Such physical quantity evaluates the width of the local velocity distribution. The size of the sliding window is chosen to suppress the contribution from the supergranules that dominate the velocity field and also from velocity structures having even larger scales (such as the differential rotation).  A sequence of the maps of mean square velocity is given in Fig.~\ref{fig:horvel}. Again, a large reproducibility of the velocity structures between to consecutive maps serves as evidence that these structures are real. Moreover, an evolution of the velocity field in the region around the filament shows some intriguing behaviour. Namely, a linear structure of an increased squared horizontal velocity in the diagonal direction crossing the middle of the filament starts to form in the panel representing the six-hour interval centred on 20 October at 21:00~UT, which prevails for around a day. It grows in magnitude, reaching its maximum in the frame indicated by the time 09:00~UT on 21 October with a significant increase of mean squared velocity near the northern end and starts to decay there after. The filament started to disperse soon after. 

The filament disintegration started rather abruptly at the northern end. Therefore, studying the fast evolution of the magnetic field is not possible by using helioseismic datasets. Thus, we used the advantage of tracking the granules using the CST code, which yielded the full-disc surface velocity field with a cadence of 30 min and an effective spatial resolution of 2.5~Mm. The pipeline provided us with the velocity components projected to the spherical coordinates, that is the zonal, meridional, and radial components. 

Over the day of the filament eruption, that is 21 October, we noticed a faster evolution only in the zonal component. To quantify these changes, we computed the longitudinal average of the zonal velocity (representing the local differential rotation profile) in the region of the filament and fitted a linear dependence with latitude to the zonal velocity profile. The slope of the linear fit indicated the average latitudinal shear in the zonal velocity component. 

\begin{figure}
    \centering
    \includegraphics[width=0.5\textwidth]{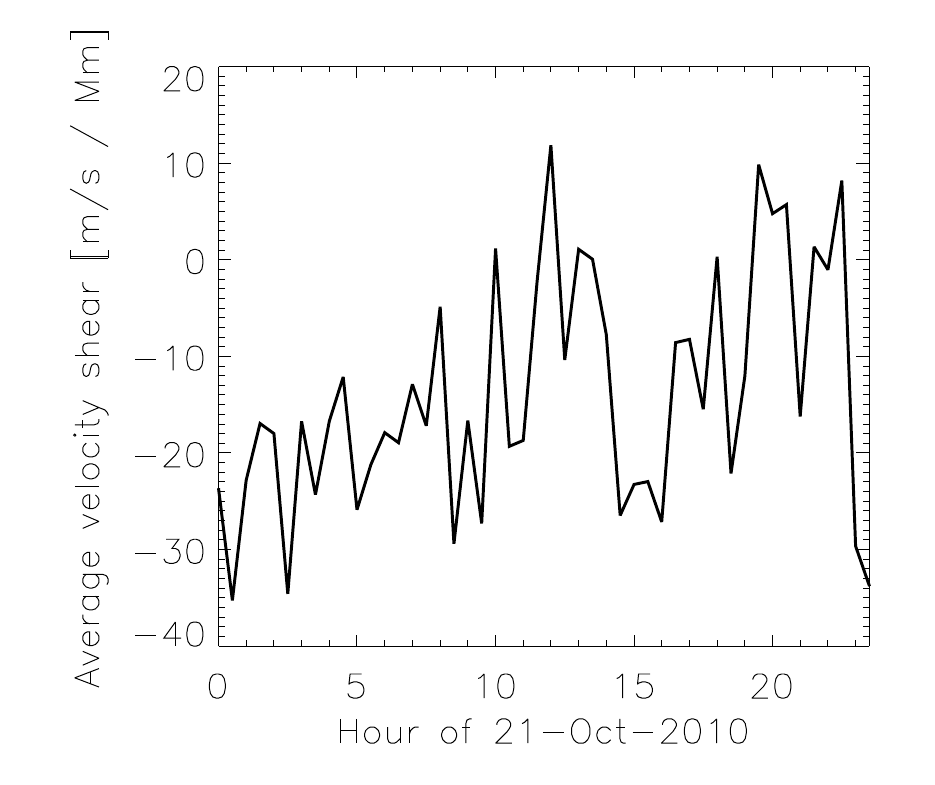}
    \caption{The evolution of the zonal shear around the filament determined from the granule tracking. }
    \label{fig:shear}
\end{figure}

It turned out (see Fig.~\ref{fig:shear}) that the average zonal shear was increasing from the beginning of the day of the eruption until about noon, when it decreased suddenly. This corresponds to a beginning of the very fast evolution of the filament seen in H$\alpha$ and 30.4-nm lines when the filament legs moved westwards very fast. The average zonal shear reached the minimum at around 15:00~UT and then increased again. The average zonal shear in the region of the former filament grew again until almost midnight on 21 October and then dropped again. By that time, the filament did not exist in H$\alpha$ and AIA 30.4~nm images anymore. 

\subsection{Evolution of the magnetic field}
The filament structure is fully locked in the structure of the chromospheric and coronal magnetic field. Unfortunately, direct measurements of this magnetic field are not available. Only measurements of the line-of-sight component of the magnetic field in photospheric layers are routinely available from SDO/HMI. We used a sequence of the 45-second magnetograms to study the temporal changes of the magnetic field in the region of the filament. 

The region of interest is located mainly in the weak-field regions, where the photon noise indicated by the random error in the magnetic field determination is significant. Thus, we averaged the line-of-sight magnetograms over one hour and sampled the series with a critical sampling rate to increase the signal-to-noise ratio. Such movies do not indicate changes that could possibly be responsible for the filament's eruption.  

To get an understanding of how the structures of the atmospheric magnetic field behave with time, we performed a magnetic-field extrapolation using the potential approximation. Potential extrapolation reconstructs only the large-scale features of the magnetic field, and by no means could it catch the details that are expected in the plasma-field coupling in the filament structures, where a detailed MHD modelling would be a proper approach. On the other hand, for instance, the motions of the filament legs observed in the eruptive phase in 30.4~nm channel of SDO/AIA are so large that is it reasonable to expect to see some changes even in the potential component of the magnetic field.  

\begin{figure*}
    \centering
    \includegraphics[width=\textwidth]{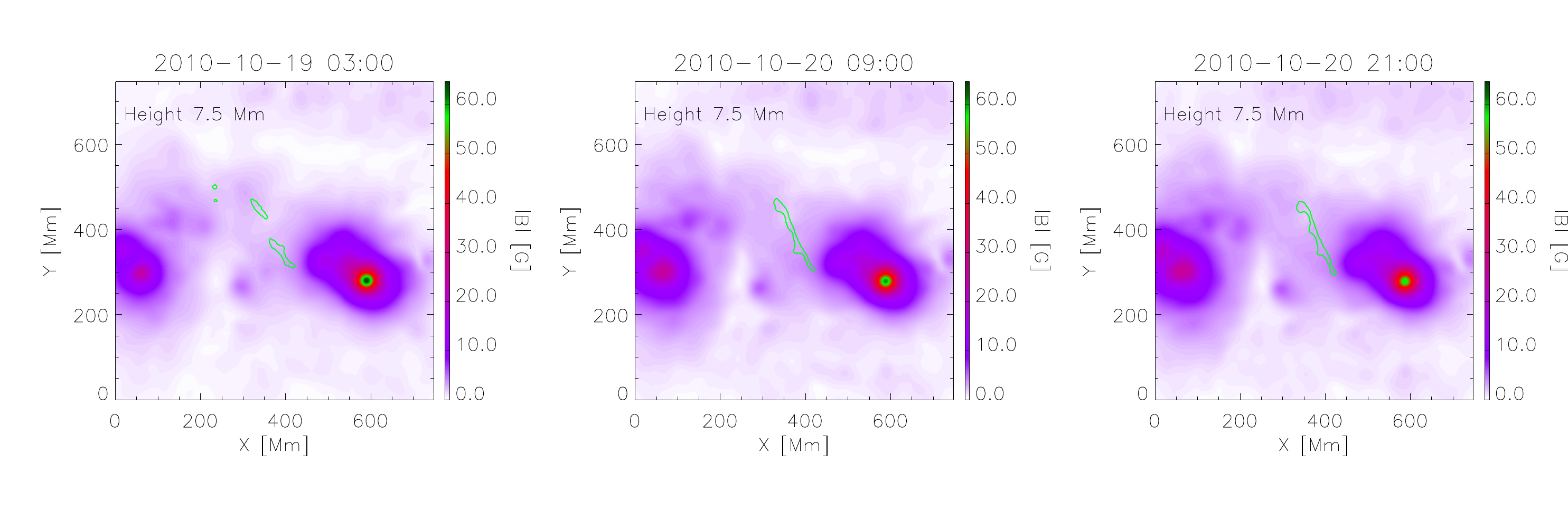}
    \includegraphics[width=\textwidth]{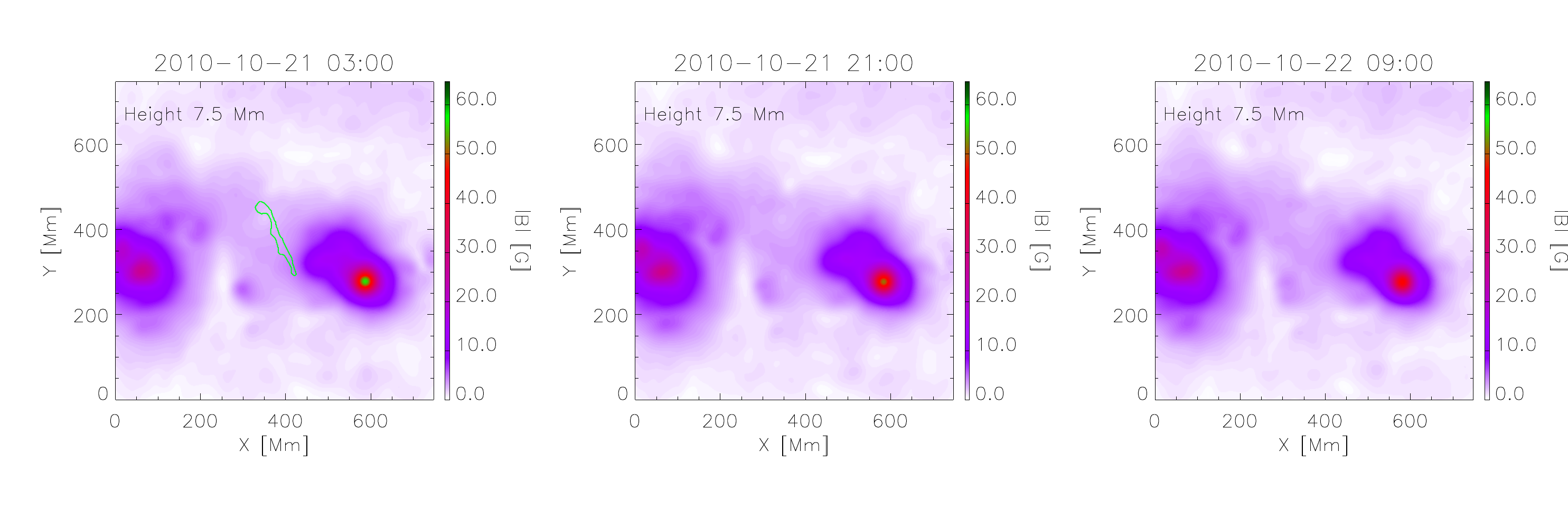}
\caption{Extrapolated total magnetic induction at a height of 7.5~Mm displayed for a few points in time. One can see a slow evolution without abrupt changes. }
    \label{fig:mg_evol50}
\end{figure*}

\begin{figure*}
    \centering
    \includegraphics[width=\textwidth]{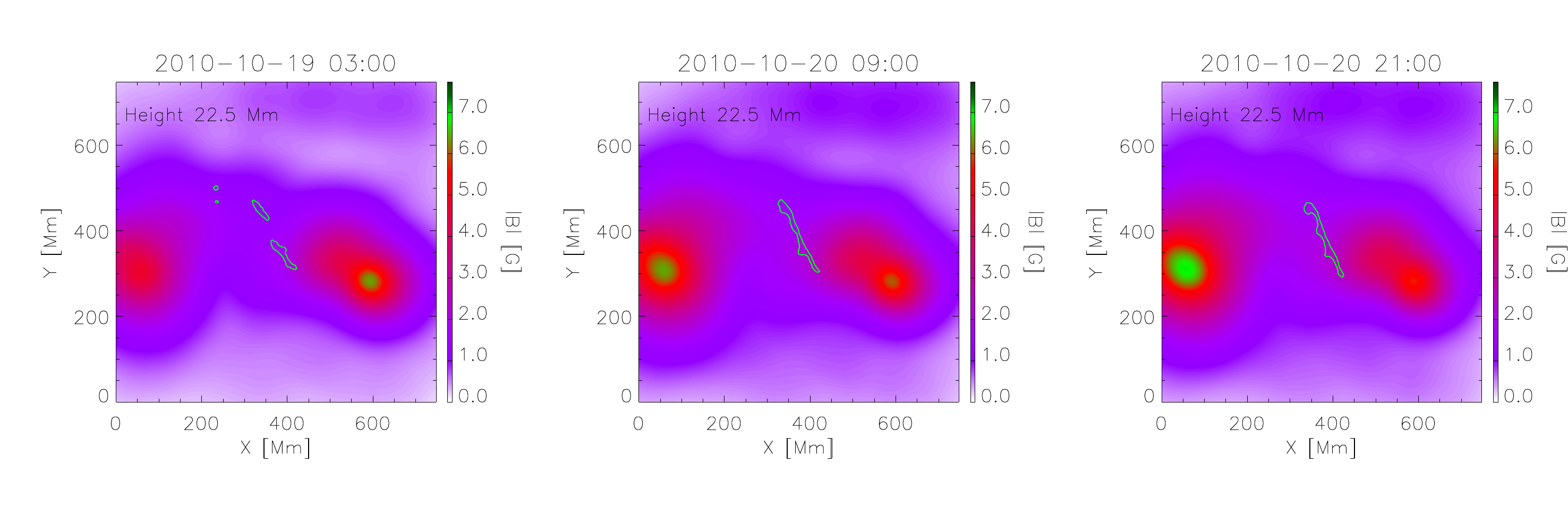}
    \includegraphics[width=\textwidth]{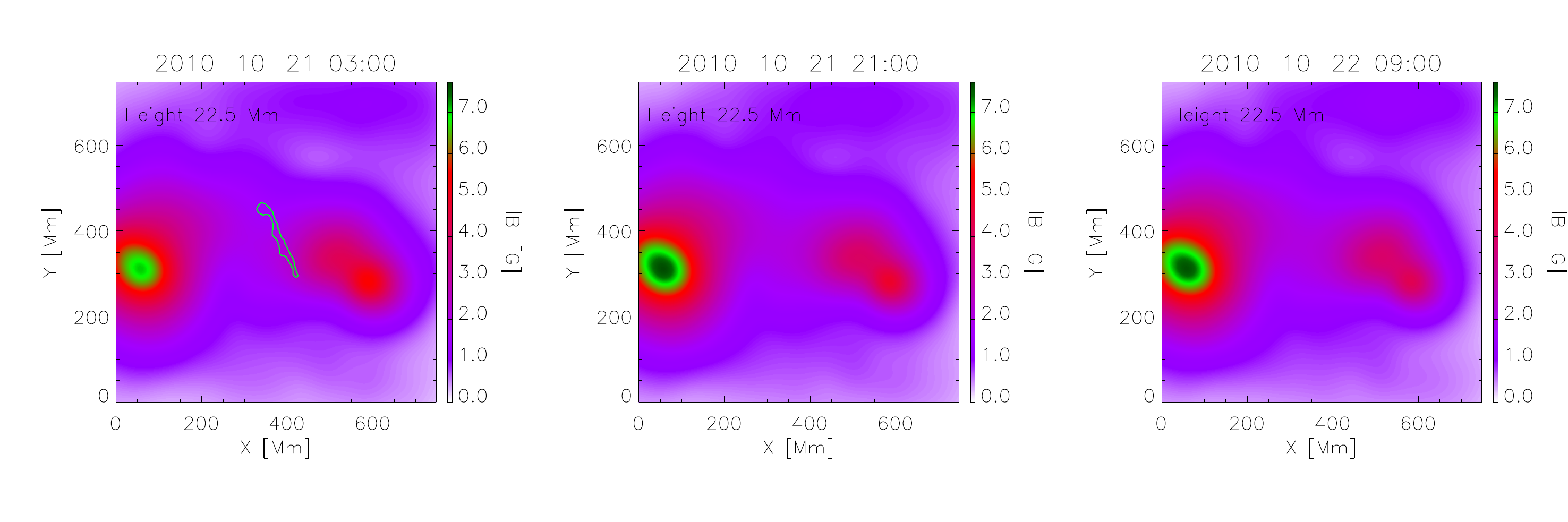}
\caption{Magnitude of the magnetic field induction at a height of 22.5~Mm. One sees changes in the evolution of the potential component of the coronal magnetic field that may have to do with the changes in the filament. Namely, the eastern part strengthens with time, whereas the western part, closer to the filament, weakens. }
    \label{fig:mg_evol150}
\end{figure*}

Our expectations were confirmed by computing the evolution of the magnetic field structures, namely in the higher atmosphere, say 10~Mm and above. While below this boundary the structures are obviously dominated by small-scale features that do not evolve significantly (see Fig.~\ref{fig:mg_evol50}), the upper field changes significantly (see Fig.~\ref{fig:mg_evol150}). There, we see the filament spine located between the two stronger polarities, closer to the western one. It is a bit stronger than the eastern polarity at the beginning of the filament evolution, that is during its increasing phase. Later, the western polarity weakens, whereas the eastern grows in intensity. Also after the filament eruption, the western part gains a different shape, it disperses a bit towards the east. 

When investigating the stability of the magnetic features, a decay index $n$ of the magnetic field, defined as $n= \der \ln B/\! \der \ln Z$, where $B$ is the induction of the magnetic field and $Z$ the height (positive from the photosphere towards corona), proves useful. Many previous studies \citep[such as][]{2016ApJ...830..132L,2017ApJ...843L...9W} showed that when $n > n_{\rm crit}$, where $n_{\rm crit}$ is usually about 1.5, the magnetic field is prone to torus instability. For instance, filaments usually erupt in the regions, where this condition is fulfilled. For the computation of the decay index, the potential extrapolation is sufficient, as the properties of the `external' magnetic field are important. 

\begin{figure*}
    \centering
    \includegraphics[width=\textwidth]{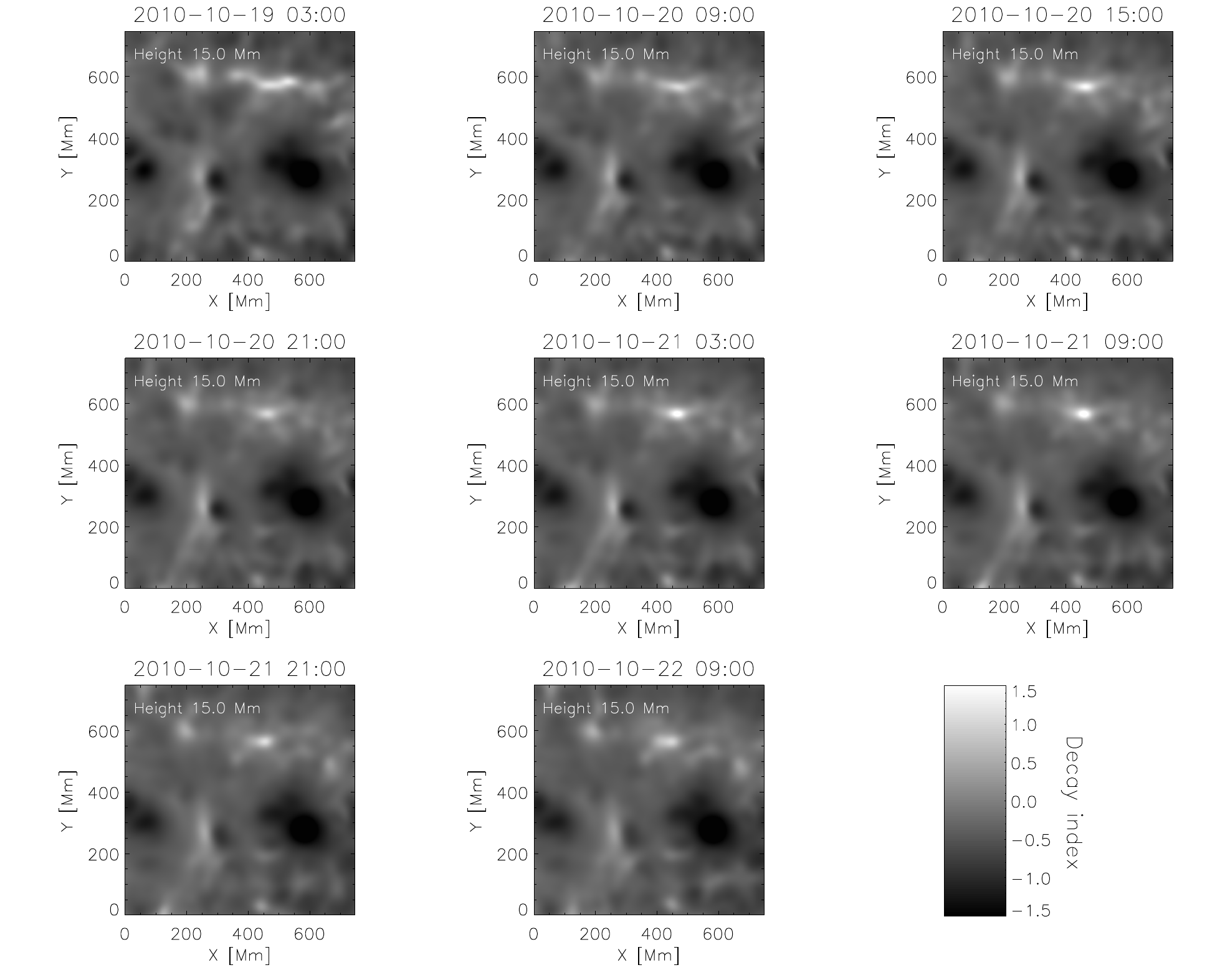}
\caption{Decay index $n$ of the magnetic field at a height of 15~Mm displayed for a set of moments in the evolution of the filament. All panels bear the same colour scale, with black indicating value $n=-1.5$ and white the value $n=+1.5$. A particular feature is visible roughly at coordinates $(440,580)$, where the filament hook is connected (compare to Fig.~\ref{fig:cuts304}). }
    \label{fig:mg_n}
\end{figure*}

In Fig.~\ref{fig:mg_n}, we give a sequence of decay-index maps at a height of 15~Mm. At positions, where the filament is located (see Fig.~\ref{fig:mg_evol150}), the value of $n$ is around zero. A significantly positive horizontal feature is seen in a row at $Y\sim550$~Mm, along which the northern footpoint of the filament seems to have slipped in the sequence of AIA 30.4 nm images. A very bright feature within this structure, which remains isolated later in the sequence, is cospatial with the anchor of the filament hook as indicated in Fig.~\ref{fig:cuts304}. The value of the decay index increases to the point immediately before the filament activation and eruption and decreases afterwards. Such an evolution is consistent with the filament activation starting at this point in the late morning of 21 October.  

\section{Conclusions}
We performed a thorough analysis of the observational data of a quiet-Sun filament, which erupted in the late afternoon on 21 October 2010. The filament eruption was not followed by an X-ray flare according to the GOES and RHESSI archives. 

The filament evolved slowly in a few days preceding the eruption, with a rapid change and disappearance on that day. We found the following:
\begin{itemize}
    \item In H$\alpha$ filtergrams, the filament area was continuously growing in the days before the eruption. The area dropped significantly as the filament activated within a few hours and the filament completely disappeared from the H$\alpha$ filtergrams in the following hours. 
    \item The filament was visible earlier in 30.4~nm AIA channel than in H$\alpha$ line-core filtergrams. It also extended to a greater length in the 30.4~nm channel. In the north, it was connected to a particular location with a large hook. As the filament activated, the filament's leg swept westward with an apparent velocity of a few kilometers per second, whereas the location of the footpoint did not change. After the disappearance of the filament, we observed a bright band in AIA 30.4~nm spectral range extending to both sides from the former location of the filament's spine. 
    \item Photospheric flows show a few peculiarities. Before the filament activation, we observed a temporal increase of the converging flows towards the filament's spine. In addition, the squared velocity increased temporarily before the activation and peaked just before it, followed by a steep decrease. In the maps of squared velocities, we see a formation of a large-velocity linear pattern, magnitude of which peaked before the filament activation. 
    \item In the flows from granular tracking, we see an increase of average shear of the zonal velocity component in the filament's region before its activation, followed by the steep decrease.
    \item The photospheric line-of-sight magnetic field shows a monotonic increase of induction eastward from the filament spine, in the region where, three days later, a new active region emerged. We performed a potential magnetic field extrapolation. The change agreeing with that at the photospheric levels is much more evident at larger heights. The decay index of the magnetic field at heights around 10~Mm shows a value larger than critical at the connecting point of the northern filament hook. The value of the decay index increased there monotonically until the filament activation. Then, it decreased sharply. 
\end{itemize}

\noindent From the observations, it is evident that the filament eruption started at its northern leg, where it connected to the northern footpoint via large hook. The location of the hook, mainly its east-west component, agrees with the region in the potential magnetic field at height larger than 10~Mm, where the decay index is large, possibly larger than critical. It was shown that those are the conditions for the torus instability or a critical loss of equilibrium, which are equivalent formulations for triggers of flux-rope eruptions \citep{2010ApJ...718.1388D,2014ApJ...789...46K}. Such an instability zone is thus prone to rapid reconfiguration of the magnetic field, causing the fast motion of the filament flux rope westwards, as seen in online AIA 30.4~nm movie. After the reorganisation the values of the decay index in the instability zone decrease. 

A strong coupling between the velocity and magnetic fields in the photosphere and near-surface layers of the convection zone probably contributed to the increase of the non-potential component of the coronal magnetic field within the filament region beyond the threshold of instability onset. Namely, the observed increase of convergent flows towards PIL about a day before the filament eruption transported magnetic elements towards PIL. We also observe an overall increase of the width of local velocity distribution in the filament region, forming peculiar aligned high-velocity features. The rate of transport of the magnetic elements before the eruption in the PIL region thus increased temporarily, likely forming small loops and arcades that further sheared, possibly twisted and reconnected, and contributed to the unstable configuration of the field in the filament flux rope. This is seen, for instance, in Fig.~\ref{fig:horvel}, where two bright spots of locally increased velocity at 09:00 on 21 October correspond to large swirls seen in Fig.~\ref{fig:rectangles} at coordinates $X' \sim 300$, $Y' \sim 350$. After reaching the instability threshold, the filament erupted. We observed this local increase of squared velocity mainly in the northern half of the filament, where also the eruption started. 

Our study shows the importance of flows in the upper layers of the convection zone for influencing the stability of the structures of the magnetic field in upper layers of the solar atmosphere. A detailed investigation and confirmation of our scenario is not possible without proper MHD modelling, which is beyond scope of this paper. Our findings are consistent with those of \cite{2018A&A...618A..43R}, who concluded that the filament destabilisation was caused by the joined action of the differential rotation shear and convergent flows towards the filament. 

\begin{acknowledgements}
M\v{S} and DK were supported by the Czech Science Foundation under grant 18-06319S. We also acknowledge the support from the institution project RVO:67985815. DK is supported by the Grant Agency of Charles University under grant No. 532217. This work was granted access to the HPC resources of CALMIP under allocation 2011-[P1115]. This study stems from the BSc. project of JW supervised by M\v{S} and DK. 
\end{acknowledgements}

%\bibliographystyle{aa}  
%\bibliography{references}

\end{document}